\documentclass[12pt]{iopart}

\usepackage{graphicx}
\usepackage[breaklinks=true,colorlinks=true,linkcolor=blue,urlcolor=blue,citecolor=blue]{hyperref}

\begin{document}

\title{Gamma-rays from the binary system containing PSR J2032+4127 during its periastron passage}

\author{W{\l}odek Bednarek, Piotr Banasi\'nski \& Julian Sitarek}
\address{University of {\L }\'{o}d\'{z}, 
Department of Astrophysics, Pomorska 149/153, 90-236 {\L }\'{o}d\'{z}, Poland}
\ead{bednar@uni.lodz.pl}
%, p.banasinski@uni.lodz.pl, jsitarek@uni.lodz.pl}

\begin{abstract}
The energetic pulsar, PSR J2032+4127, has been recently discovered in the direction of the unidentified HEGRA TeV $\gamma$-ray source (TeV J2032+4130). It is proposed that this pulsar forms a binary system with the Be type star, MT91 213, expected to reach the periastron late in 2017. 
We performed detailed calculations of the $\gamma$-ray emission produced close to the binary system's periastron passage by applying a simple geometrical model. Electrons accelerated at the collision region of  pulsar and stellar winds initiate anisotropic Inverse Compton $e^\pm$ pair cascades by
scattering soft radiation from the massive companion. The $\gamma$-ray spectra, from such a comptonization process, are confronted with the measurements of the extended TeV $\gamma$-ray emission from the HEGRA TeV $\gamma$-ray source. 
We discuss conditions within the binary system, at the periastron passage of the pulsar, for which the $\gamma$-ray emission from the binary can overcome the extended, steady TeV $\gamma$-ray emission from the HEGRA TeV $\gamma$-ray source. 
\end{abstract}

%Uncomment for PACS numbers title message
%\pacs{00.00, 20.00, 42.10}
%%??\pacs{95.55.Ka;95.55.Vj;95.75-z;95.75.Mn;95.85.Pw;95.85.Ry}

% Keywords required only for MST, PB, PMB, PM, JOA, JOB? 
%\vspace{2pc}

% Uncomment for Submitted to journal title message

\submitto{\JPG}

% Comment out if separate title page not required
\maketitle

\section{Introduction}

Binary systems, containing a young, energetic pulsar and a massive companion star, provide the unique opportunity for the
investigation of high energy processes occurring within the pulsar winds and their termination shocks.
Relativistic particles, accelerated in the pulsar vicinity, find a well defined target in the stellar wind and the soft radiation from a companion star. Up to now, GeV-TeV $\gamma$-ray emission from only one such a system has been observed and studied in details at the TeV energies \cite {aha05a,aha09,abr13} and at the GeV energies \cite{ab11,tam11,tam15}. It contains radio pulsar PSR B1259-63 and Be type massive companion LS2883.
 A few other TeV $\gamma$-ray binaries are also suspected to harbour young pulsars. However, due to the compactness of their systems, the nature of the compact objects remains unknown. The details of the observations of the TeV $\gamma$-ray binaries, and their theoretical interpretation, have been recently reviewed in \cite{dub13}.

Another young pulsar, PSR J2032+4127, has been proposed to be the companion of a massive Be type star, MT91 213. Its parameters are similar to those of the star LS2883 \cite{lyn15}. This binary system is much more extended than PSR B1259-63/LS2883. PSR J2032+4127 is expected to reach the periastron late in 2017 \cite{ho17}. 
The high energy X-ray and GeV $\gamma$-ray emission from this binary has been recently studied in 
\cite{tak17, bir17}. It is found that the synchrotron X-ray emission increases rapidly but the GeV emission remains constant when the pulsar is approaching the periastron.  The observations of this binary system in the TeV $\gamma$-ray energy range are expected to be more difficult since an extended, steady TeV $\gamma$-ray source (TeV J2032+4130) has been discovered by the HEGRA Collaboration in the direction of this binary \cite{aha02} (and confirmed in \cite{alb08, ali14}). 
%{\bf No evidence of an enhanced TeV $\gamma$-ray emission from this binary system before the summer 2017 has been seen \cite{bir17}. 
Moreover, in September 2017, for the first time such an enhanced TeV $\gamma$-ray emission has been reported by the VERITAS and MAGIC Collaborations \cite{mv17}.

We discuss the Inverse Compton (IC) e$^\pm$ pair cascade model in the context of the binary system PSR J2032+4127/MT91 213. We present detailed calculations of the GeV-TeV $\gamma$-ray radiation from this binary close to the periastron passage of the pulsar applying the available parameters of PSR J2032+4127/MT91 213.
The Monte Carlo cascade code, which follows the production of $\gamma$-rays by relativistic electrons accelerated in the wind collision region, has been used. The electrons comptonize a well defined radiation field from the companion star. The results of the calculations are confronted with the observations of the TeV $\gamma$-ray emission from 
the HEGRA source, TeV J2032+4130. Our aim is to determine the conditions, close to the periastron passage of PSR J2032+4127/MT91 213, for which the TeV $\gamma$-rays produced within the binary system can be observed over the steady TeV $\gamma$-ray emission from the HEGRA TeV source.

\section{The binary system PSR J2032+4127/MT91 213}

The pulsed GeV emission, with the period of $P = 143$ ms, has been discovered from the pulsar PSR J2032+4127 by the $\it{Fermi}$-LAT \cite{ab09}.
The spin down power of this pulsar is $L_{\rm PSR} =  2.7\times 10^{35}$ erg~s$^{-1}$, the characteristic age is $115.8$ kyr, and the surface magnetic field is estimated on $B_{\rm PSR} = 1.7\times 10^{12}$ Gs \cite{ab09} (see also radio detection \cite{cam09}). The pulsar is on an elongated orbit around the companion star, MT91 213, which is a B0 V type star with the luminosity $L_\star = 5.8\times 10^{37}$~erg~s$^{-1}$ \cite{wri15}. This luminosity corresponds to a star with the effective surface temperature, the radius, and the distance as given in Table~1 \cite{wri15}. Note that the radius and the luminosity of the star are lower than those used in \cite{tak17} by a factor of 7/3 and $\sim5.4$, respectively. The luminosity of the companion star which we consider is consistent with the observations of MT91 213 \cite{wri15}. We use above parameters of the star and the pulsar in our further modelling of the radiation processes.  

\begin{table}
  \caption{The basic parameters of the binary system PSR J2032+4127/MT91 213 and the massive companion star MT91 213.}
  \begin{tabular}{llllllllllll} 
\hline 
binary \cite{ho17} & $a = 2.7\times 10^{14}$ cm  &  $e = 0.961$ & $\omega = 40^\circ$   \\
\hline
star \cite{wri15}  &  $T_\star = 3.1\times 10^4$ K   &  $R_\star = 3\times 10^{11}$ cm & $d = 1.33\pm 0.06$ kpc \cite{kim15} \\
\hline 
\end{tabular}
  \label{tab1}
\end{table}

The basic parameters of the pulsar orbit in the binary system have been recently updated. However, they are still not precisely known since only a part of the pulsar orbit has been observed so far. According to the model 2 of \cite{ho17}, the values of the semi-major axis of the binary system, its eccentricity and the longitude of the periastron are given in Table 1.
The distance of the pulsar from the companion star, during the periastron passage, is then  
$D_{\rm peri} = a(1 - e)\approx 1.05\times 10^{13}$ cm. The distance between the stars at the superior conjunction of the pulsar is $D_{\rm sup} =  a(1 - e^2)/[1 + e\cos(90^\circ - \omega)]\approx 1.28\times 10^{13}$ cm, where the longitude of periastron is $\omega = 40^\circ$ 
(again, model 2 in \cite{ho17}). The angle $\alpha$, between the line of sight and the direction defined by the centres of the stars at the periastron, is $\cos(180^\circ - \alpha) = \cos(90^\circ + \omega) \cos(90^\circ - i)$, where $i$ is the inclination angle of the binary system (see Fig.~1). 
$\alpha$ is equal to $90^\circ$ for the inclination of the binary $i = 0^\circ$ and $\sim 56^\circ$ for $i = 60^\circ$. 

\begin{figure}
%\centering
\vskip 6.5truecm
\includegraphics{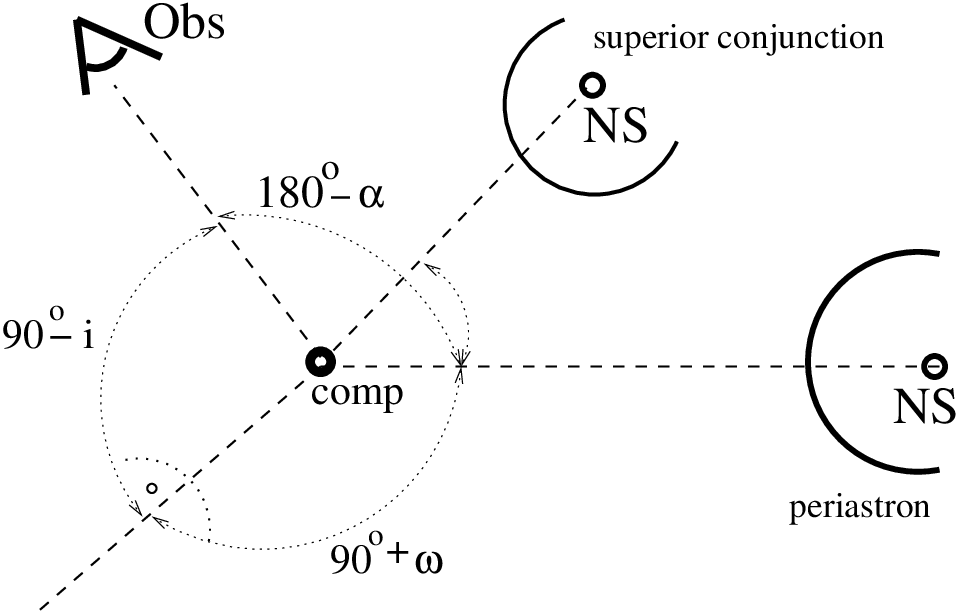}
\caption{Schematic representation of the geometry of the binary system containing PSR J2032+4127 at the periastron passage and at the superior conjunction.
The pulsar is surrounded by the wind collision region (thick solid curve), in which electrons are accelerated. The observer sees the companion star at the angle $\alpha$, which is measured between the directions defined by the centres of the companion star "comp" and the pulsar "NS" and the direction towards the observer "Obs". The inclination angle of the binary system is $i$ and the longitude of the periastron passage of the pulsar is $\omega$.}
\label{fig1}
\end{figure}

The pulsar PSR J2032+4127 is expected to form a pulsar wind. Its interaction with the wind of the companion star
can be quite complicated. Far from the periastron, the mixed pulsar and stellar winds can form a nebula around the pulsar (e.g. \cite{bb11}). Colliding pulsar and stellar winds form a spiral, double-armed structure (e.g. \cite{lam12}). In the inner part of the nebula
particles can be accelerated in the region of colliding winds producing high energy radiation \cite{bs13,zab13}. Close to the periastron, the winds create a collision region between the star and the pulsar in which particles can be effectively accelerated. If the winds are homogeneous, then large scale 
shock waves can be formed.  Particles can be accelerated to energies allowing them  production of X-ray radiation
in the synchrotron process and TeV $\gamma$-rays in the Inverse Compton (IC) scattering of the stellar radiation \cite{mt81,tak94,ta97,kir99}. 
The fate of the pulsar wind after the collision with the stellar wind is not well known. 
Some models of the collision of homogeneous winds within the binary system predict that the two winds form separate fluids which do not mix efficiently. In such a case, the pulsar wind can even accelerate along the shock structure as a result of the adiabatic expansion \cite{bog08,dub10,kon12}. 
However, homogeneous winds can also mix efficiently due to the instabilities which appear during wind collisions as shown in two and three dimensional hydrodynamical simulations \cite{br12,br15}.
In the case of very inhomogeneous winds, the stellar wind can load the pulsar wind with the barionic matter \cite{bed14,par15,de17}.
Then, both winds mix to some extent. If the winds mix very effectively, then the pulsar wind can be significantly decelerated to the velocity $v_{\rm mix}$. 
We make an order of magnitude estimate of the winds velocity after mixing by assuming a simplified scenario in which the colliding winds conserve their energy and momentum. The conservation of the wind energy gives us the upper limit on the velocity of the mixed winds since a part of the energy is expected to be thermalised in the collision process.
Then, assuming that the pulsar wind is being loaded with the matter from the stellar wind \cite{bed14}, the velocity of the mixed winds after the shock can be roughly estimated by 
\begin{eqnarray}
0.5L_{\rm PSR}\approx 0.5\Delta\Omega \dot{M}_\star v_{\rm mix}^2,
\label{eq1}
\end{eqnarray}
\noindent
where $v_{\rm mix}$ is the velocity of the winds after mixing, $\Delta\Omega\approx (R_{\rm sh/PSR}/D)^2/4\approx \eta/4(1 + \sqrt{\eta})^2$ is the solid angle of the shock as seen from the centre of the companion star, and $D = 10^{13}D_{13}$ cm is the separation of the stars, $\eta = L_{\rm PSR}/(c\dot{M}_\star v_\star)\approx 0.14/M_{-8}v_8$ is the pressure ratio of the winds from the sites of pulsar and the companion star \cite{gw87}, $\dot{M}_\star = 10^{-8}M_{-8}$ M$_\odot$~yr$^{-1}$ and $v_\star = 10^8v_8$ cm~s$^{-1}$ are the mass loss rate and the velocity of the wind, $L_{\rm PSR} = 10^{35}L_{35}$ erg s$^{-1}$ is the pulsar power, and $c$ is the speed of light.
The mixed wind velocity is then,
\begin{eqnarray}
v_{\rm mix}\approx 3.5\times 10^9(1 + \sqrt{\eta})v_8^{1/2}~~~{\rm cm~s^{-1}}.
\label{eq2}
\end{eqnarray}
\noindent
Note that in such a case, electrons accelerated in turbulent winds are relatively slowly advected from the
vicinity of the companion star. This provides good conditions for  more efficiently comptonize the stellar radiation to the $\gamma$-ray energy range. In this limit, 
the velocities of the mixed winds can reach a few percent of the speed of light. 
On the other hand, the simple conservation of momentum of the colliding laminar winds gives the lower limit on the mixed wind velocity \cite{crw96}. 
In this case, the mixed wind velocity is close to zero at the apex. It increases up to the velocity of the stellar wind 
(before mixing) at the parts of the collision region which are far away from the apex. In our calculations we use the range of velocities which are laying between those 
two limits. 

The very turbulent pulsar wind, loaded with the matter of the stellar wind, moves along the collision region of the winds. The distance, between the apex of the wind collision region and the pulsar, is estimated from, 
\begin{eqnarray}
R_{\rm sh/PSR} = D\sqrt{\eta}/(1 + \sqrt{\eta}).
\label{eq3}
\end{eqnarray}

\section{Gamma-ray production}

We assume that electrons are accelerated in the magnetized, turbulent collision region of the barion loaded  pulsar wind. The acceleration time scale is determined by the acceleration coefficient $\xi$ and the magnetic field strength
\begin{eqnarray}
\tau_{\rm acc} = {{R_{\rm L}}\over{\xi c}}\approx {{10E_{\rm TeV}}\over{\xi_{-2}B_{\rm sh}}}\approx
{{91 E_{\rm TeV}\sqrt{\eta}D_{13}}\over{\xi_{-2}\sigma_{-1}^{1/2}(1 + \sqrt{\eta})}}~~~{\rm s},
\label{eq5}
\end{eqnarray}
\noindent
where $R_{\rm L}$ is the Larmor radius of electrons with energy $E_{\rm e} = 1E_{\rm TeV}$ TeV, $B_{\rm sh}$ is the magnetic field strength at the shock, and $\xi = 0.01\xi_{-2}$.
These electrons are also advected with the velocity $v_{\rm mix}$ from the acceleration site. 
The characteristic time scale for the advection process can be estimated from,
\begin{eqnarray}
\tau_{\rm adv}\approx R_{\rm sh/PSR}/v_{\rm mix}\approx 10^4D_{13}\sqrt{\eta}/(1 + \sqrt{\eta})/v_9~~~{\rm s},
\label{eq2}
\end{eqnarray}
\noindent
where $v_{\rm mix} = 10^9v_9$ cm s$^{-1}$.
The advection time scale of the electrons limits their maximum energies,
\begin{eqnarray}
E_{\rm max}^{\rm adv}\approx 110\sigma_{-1}^{1/2}\xi_{-2}/v_9~~~{\rm TeV}.
\label{eq3}
\end{eqnarray}
\noindent
$\sigma = 0.1\sigma_{-1}$ is the magnetization parameter of the pulsar wind.
The value of $\xi$ can be related to the velocity
of the mixed winds in the following way $\xi\sim (v_{\rm mix}/c)^2$ \cite{md01}, if particles are energized after the mixing process of the winds. However, if the acceleration process of electrons occurs already in the region of the pulsar wind then, $\xi$ can be fixed to $\sim$0.1 since the pulsar wind velocity slows down to $\sim$0.3c. 

The acceleration process can be also saturated by the electron's energy losses on the synchrotron process 
(the cooling time scale is given by Eq.~8 in \cite{bed14}). Then, the maximum energies of the electrons are limited to
\begin{eqnarray}
E_{\rm max}^{\rm syn}\approx 6.1\left({{\xi_{-2}}\over{B_{\rm sh}}}\right)^{1/2}~~~{\rm TeV}
\approx 17.5\left({{\xi_{-2}\sqrt{\eta}D_{13}}\over{\sigma_{-1}^{1/2}(1 + \sqrt{\eta})}}\right)^{1/2}~~~{\rm TeV}.
\label{eq4}
\end{eqnarray}
The magnetic field at the collision region of the winds is estimated by a simple extrapolation from the vicinity of the pulsar,
\begin{eqnarray}
B_{\rm sh}\approx B_{\rm PSR} \left({{R_{\rm NS}}\over{R_{\rm LC}}}\right)^3\left({{R_{\rm LC}}\over{R_{\rm sh}}}\right)
\sigma^{1/2}\approx {{0.11\sigma_{-1}^{1/2}(1 + \sqrt{\eta})}\over{\sqrt{\eta}D_{13}}}~~~{\rm G},
\label{eq5}
\end{eqnarray}
\noindent
where $R_{\rm NS} = 10^6$ cm is the radius of the neutron star, 
$R_{\rm LC} = cP/2\pi\approx 6.8\times 10^8$ cm is the light cylinder radius of the pulsar.

\begin{figure*}
%\centering
\vskip 5.5truecm
\includegraphics{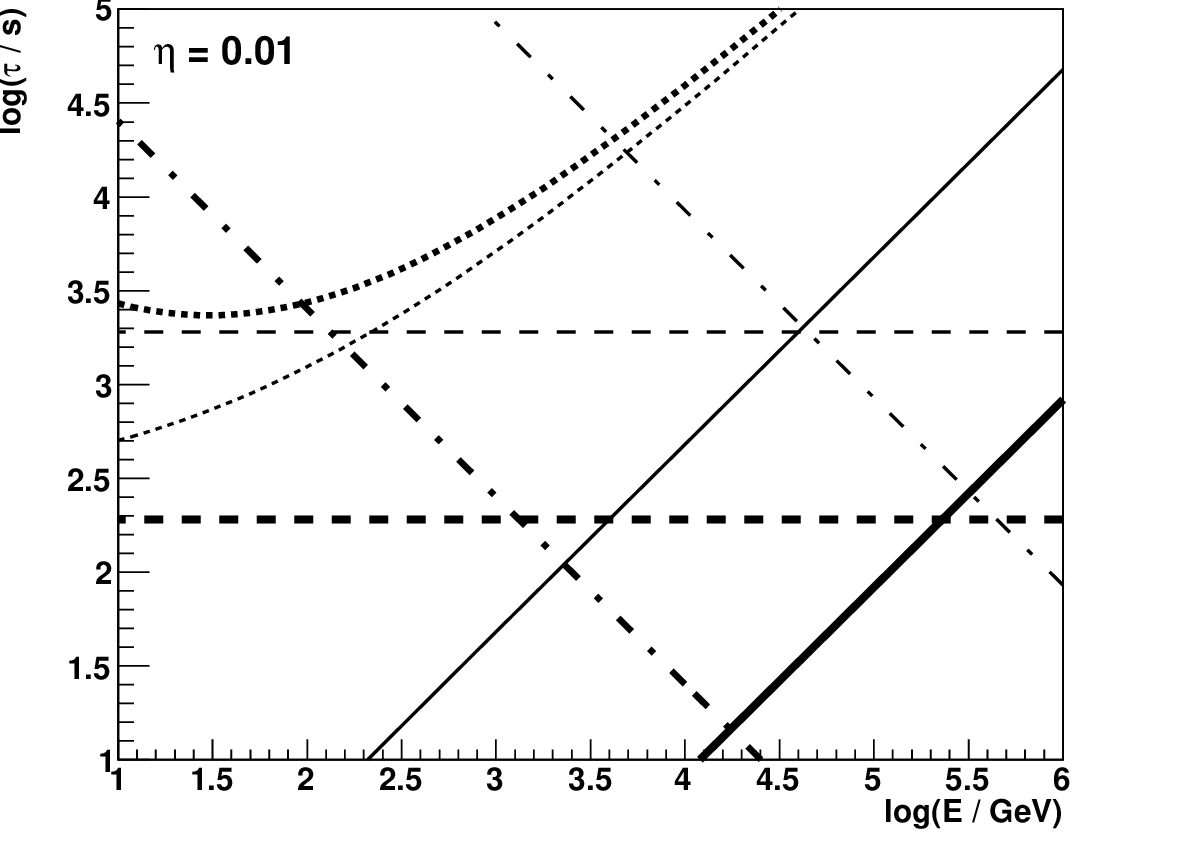}
\includegraphics{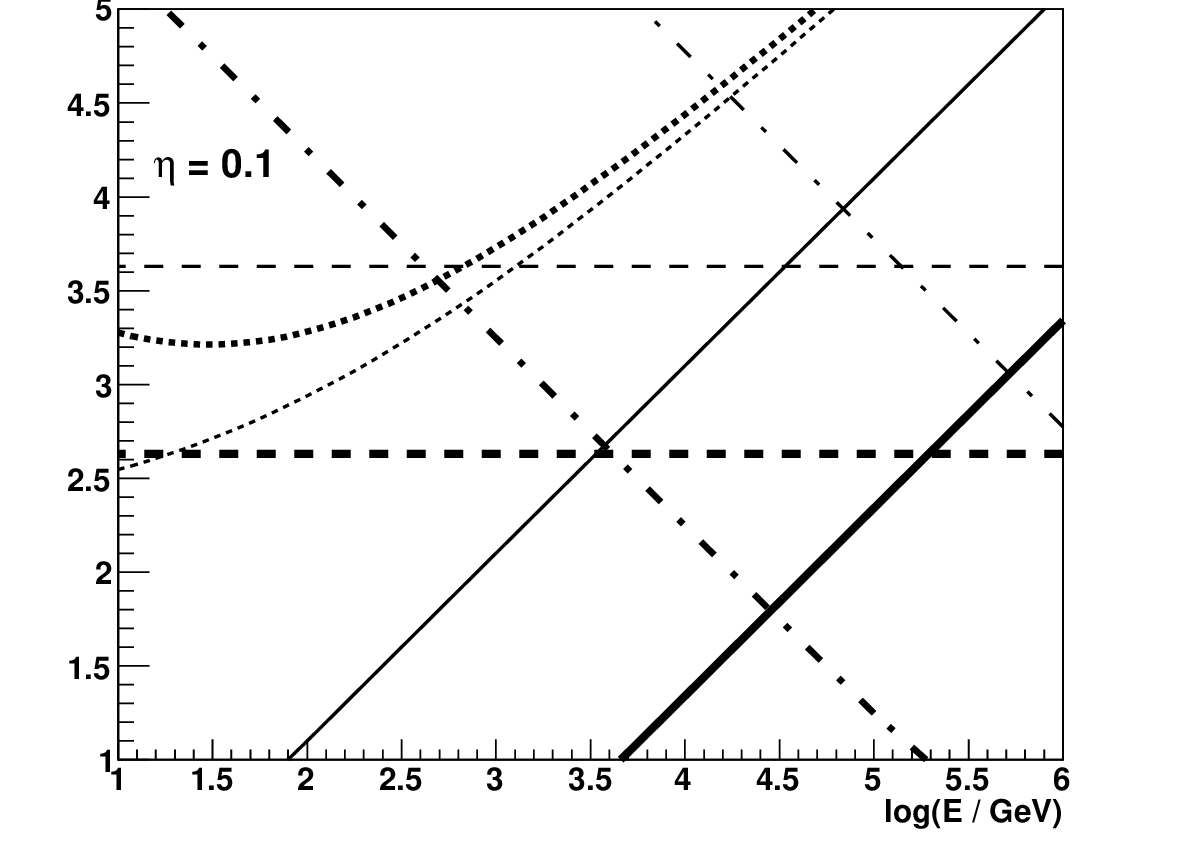}
\includegraphics{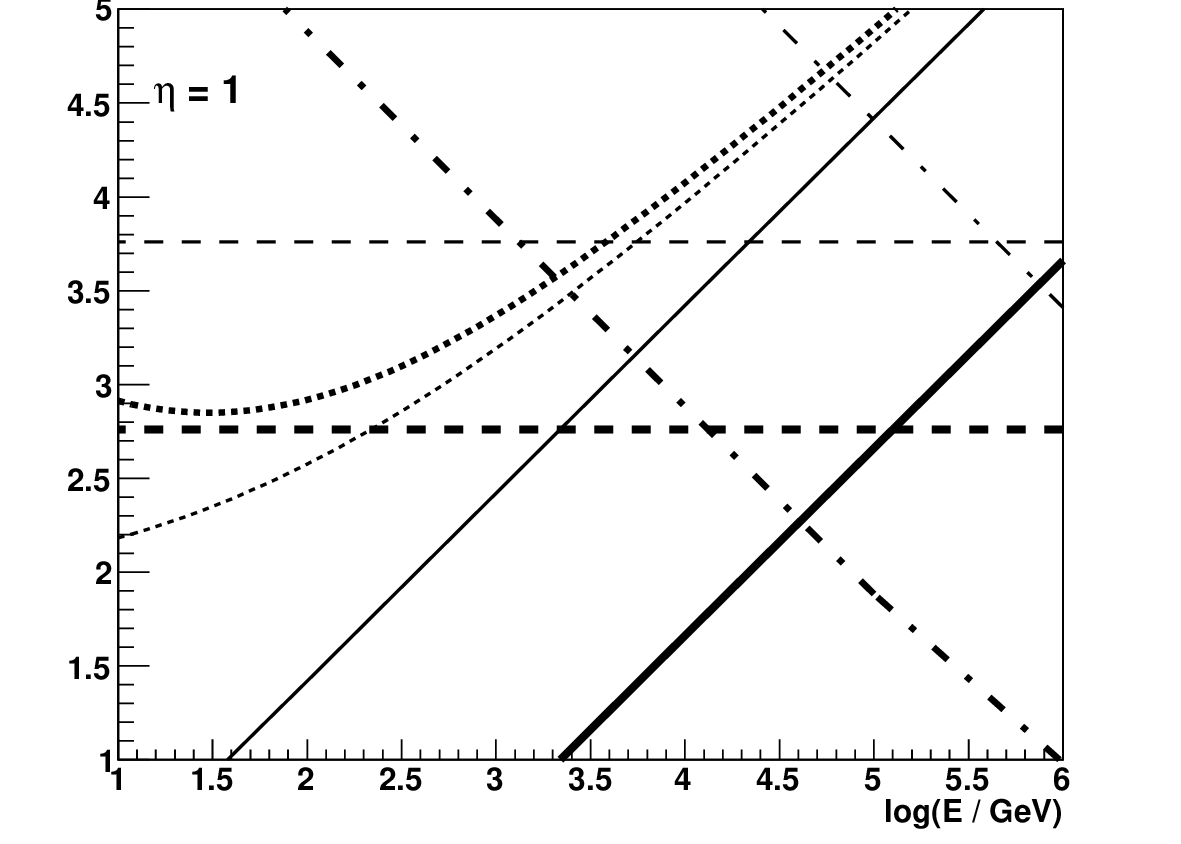}
\caption{The energy-dependent time scales, $\tau$, for relativistic electrons, which are accelerated at the wind collision region close to the periastron passage of the binary system containing PSR 2032+4127. 
Separate figures show the results for different distances of the apex of the collision region from the pulsar,
$\eta = 0.01$ (on the left), 0.1 (in the middle), and 1. (on the right). 
The acceleration time scale (solid lines), the advection time scale along the shock (dashed), the synchrotron energy loss time scale (dot-dashed) and the IC energy loss time scale in the radiation of the companion star (thick dotted) and the IC scattering time scale (thin dotted). The acceleration time scale is shown for the product of the acceleration parameter and the magnetization parameter of the pulsar wind $\xi \sigma^{1/2} =  0.03$ (thick solid line) and for  $\xi \sigma^{1/2} = 5.4\times 10^{-4}$ (thin solid line). The advection time scale is shown for the velocity of the stellar wind equal to $v_\star = 3\times 10^8$ cm s$^{-1}$ (thick dashed) or $3\times 10^6$ cm s$^{-1}$ (thin dashed). The synchrotron time scale is shown for $\sigma = 0.1$ (thick dot-dashed line) and  $\sigma = 0.003$ (thin dot-dashed line).}
\label{fig2}
\end{figure*}

The characteristic time scales for the electrons, accelerated at the collision region of the winds (i.e. for the acceleration process, the advection from the apex of the shock, the synchrotron energy losses and the IC process), are compared in Fig.~2. They depend on different parameters describing the scenario, i.e. the pressure ratio $\eta$, the magnetization parameter of the pulsar wind $\sigma$, the acceleration parameter $\xi$, and the velocity of the stellar wind $v_\star$. Fig.~2 allows us to determine the process which defines the maximum energies of the accelerated electrons at specific conditions. The electrons can be accelerated in these conditions up to several TeV.
Moreover, Fig.~2 allows us to identify the main energy loss process of electrons with different energies. It is clear that the most energetic electrons lose energy mainly on the synchrotron process. The transfer of electron's energy to $\gamma$-rays becomes efficient at energies lower than $\sim$TeV.

The flow of the mixed winds at larger distances can become very complicated as shown e.g. in \cite{ba16}.
We consider only the inner part of the collision region comparable to the dimension $R_{\rm sh/PSR}$ on which the electrons are advected on the time scale, $\tau_{\rm adv}$.  The distant part is not considered in our calculations. We assume that electrons are accelerated at the apex of the collision region of stellar and pulsar winds.
The electrons obtain a power law spectrum with the spectral index equal to -2 up to the maximum energy determined by the most constraining condition given by Eq.~6 or Eq.~7. The maximum  energies of the electrons, for specific parameters of the model, can be found on Fig. 2. Moreover, it is assumed that the relativistic electrons take $10\%$ of the spin down power of the pulsar. The angular distribution of the primary electrons is isotropic
in the plasma reference frame. The electrons are advected from the acceleration place on the advection time scale (see Eq.~5). During this escape process, the electrons lose energy on the synchrotron radiation and on the IC scattering of the anisotropic radiation from the companion star. We calculate the synchrotron spectra produced by these primary electrons. They are emitted isotropically since the magnetic field is assumed to be random in the reference frame of the plasma. In contrast, $\gamma$-rays in the IC process are produced 
anisotropically due to the anisotropic radiation of the companion star as seen from the injection place of the electrons. $\gamma$-rays, mainly produced in the general direction 
towards the companion star, are additionally absorbed in the stellar radiation. 
In order to show the importance of the $\gamma$-ray absorption effects, we calculate the optical depths for $\gamma$-ray photons with different energies for selected distances of the injection place from the star and the specific observation angles (see Fig.~3). The optical depths are clearly above unity within the specific solid cone around the star.
Due to the absorption of the primary $\gamma$-rays, the anisotropic IC $e^\pm$ pair cascade can develop in the surrounding of the massive star. In general, such type of an anisotropic cascade process can be very complicated since the magnetic fields can re-distribute the directions of the secondary $e^\pm$ pairs (see e.g. discussion of different scenarios in \cite{bed11,dub13}). In order to calculate the $\gamma$-ray spectra emerging from the binary system, we apply the specific IC $e^\pm$ pair cascade model in which the secondary 
$e^\pm$ pairs are isotropised by the random component of the magnetic field close to their place of creation (at first discussed in \cite{bed97,bed00}). 
The synchrotron energy losses of secondary $e^\pm$ pairs are not taken into account in these calculations. In fact, leptons can be isotropised by relatively weak magnetic fields which are
not able to strongly effect their energies in synchrotron process (see conditions given by Eq. 4 and Eq. 6 in \cite{bed97}).

\begin{figure*}
%\centering
\vskip 11.truecm
\includegraphics{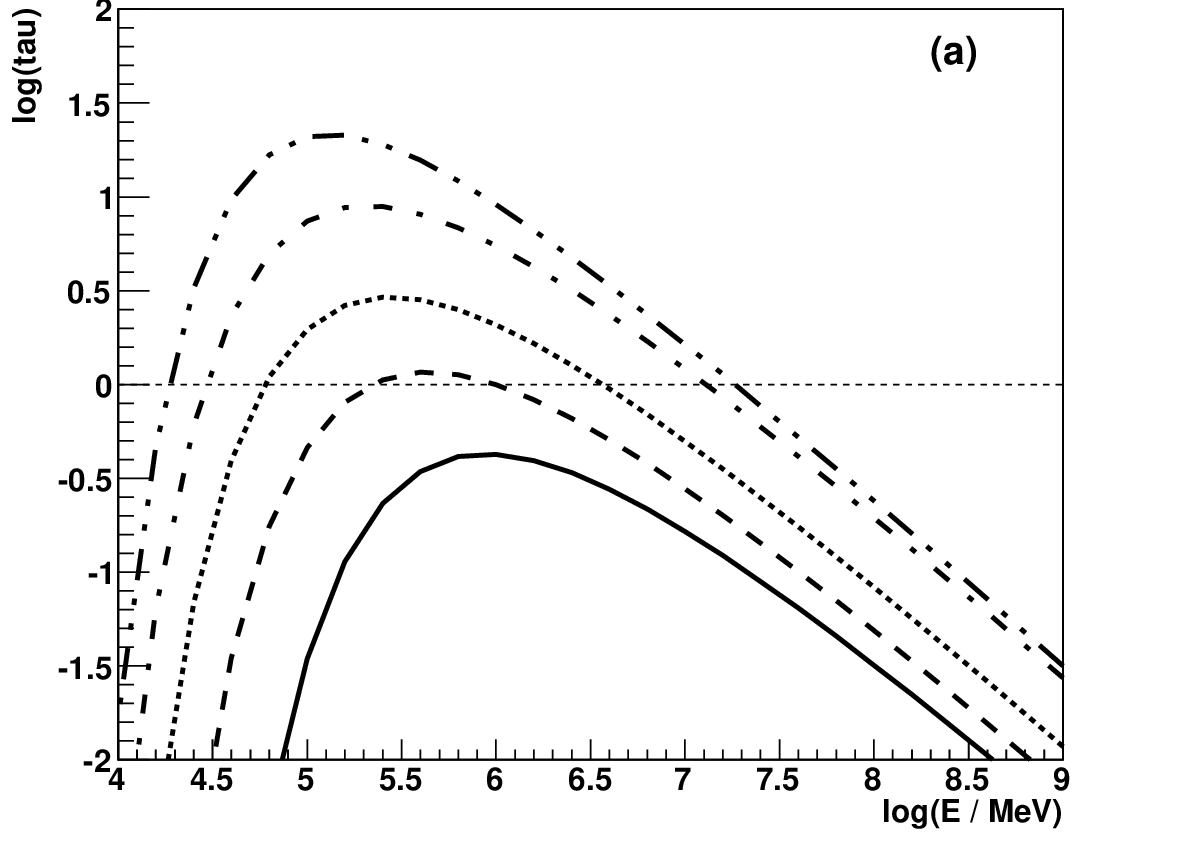}
\includegraphics{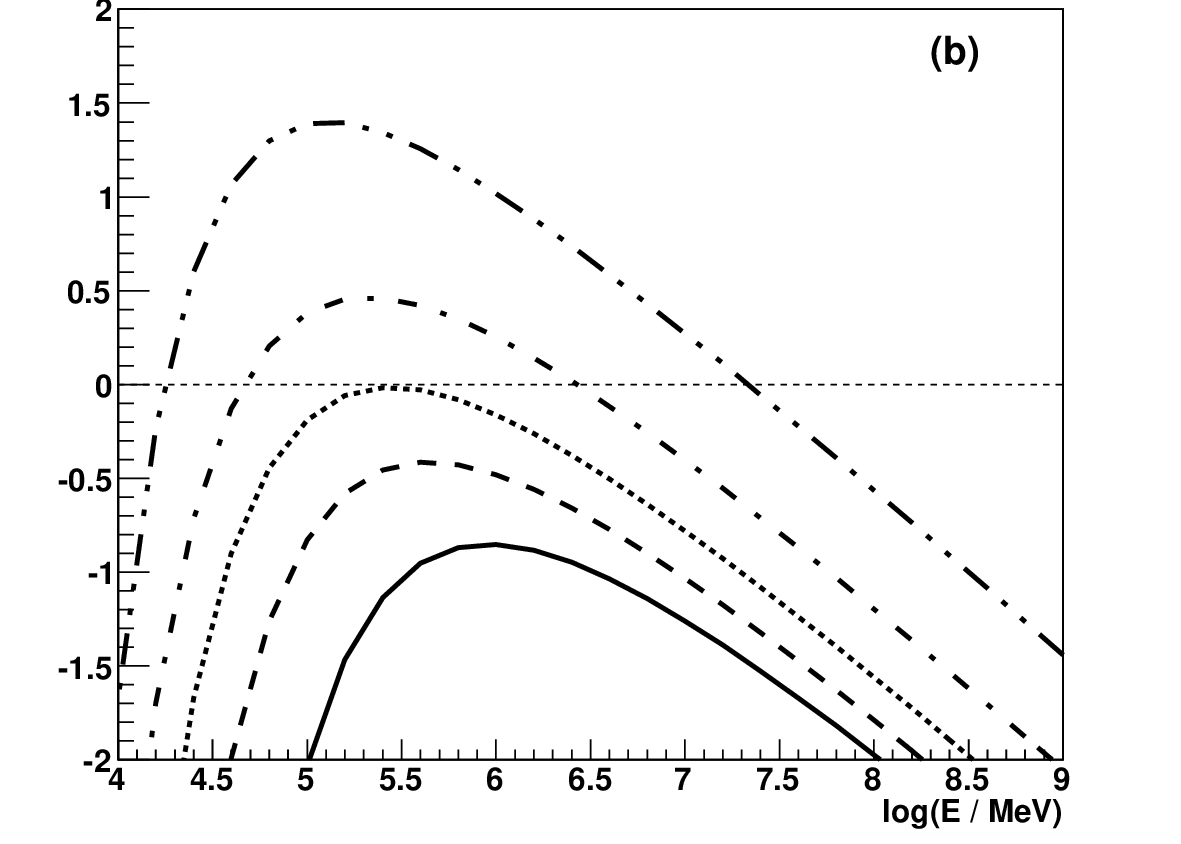}
\includegraphics{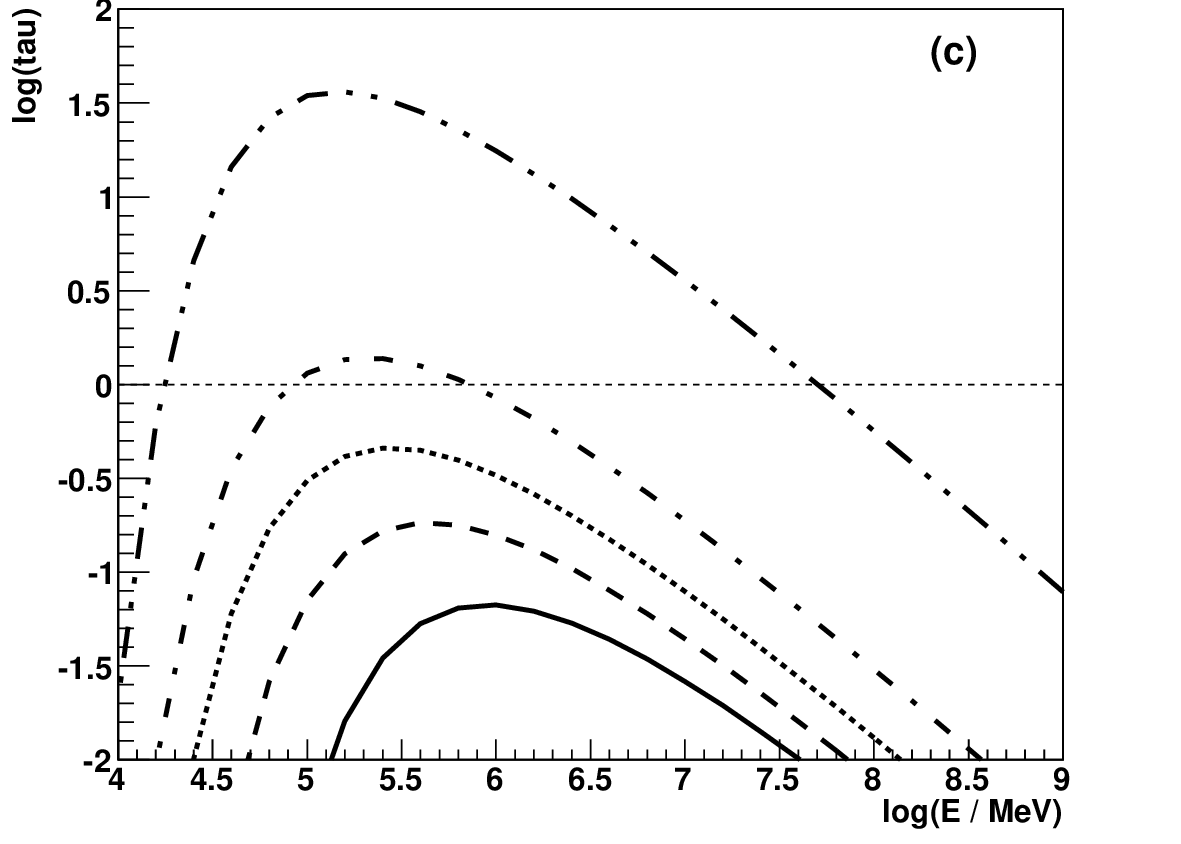}
\includegraphics{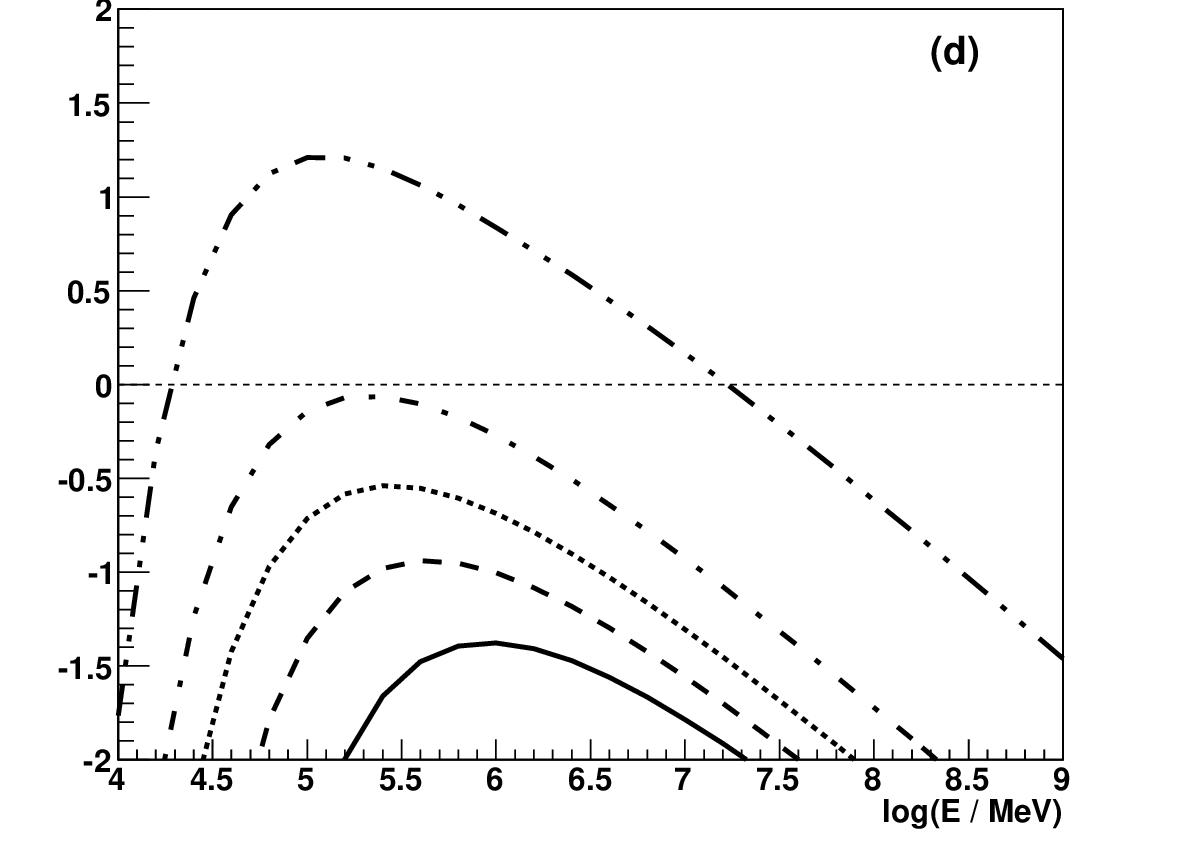}
\caption{Optical depths, "tau", for $\gamma$ rays in the radiation field of the companion star, MT91 213. 
$\gamma$ rays are injected at different distances from the centre of the 
companion star equal to $D = 5R_\star$ (figure (a)), $15R_\star$ (b), $25R_\star$ (c), and $40R_\star$ (d). 
They are injected at the angle $\alpha$, measured in respect to the direction defined by the stars, $\alpha = 0^\circ$ (dot-dot-dashed curve), $30^\circ$ (dot-dashed), $60^\circ$ (dotted), $90^\circ$ (dashed), and $120^\circ$ (solid). The star has the surface temperature $T_\star = 3.1\times 10^4$ K and the radius $R_\star = 3\times 10^{11}$~cm.}
\label{fig3}
\end{figure*}
\begin{figure}
\vskip 5.5truecm
\includegraphics{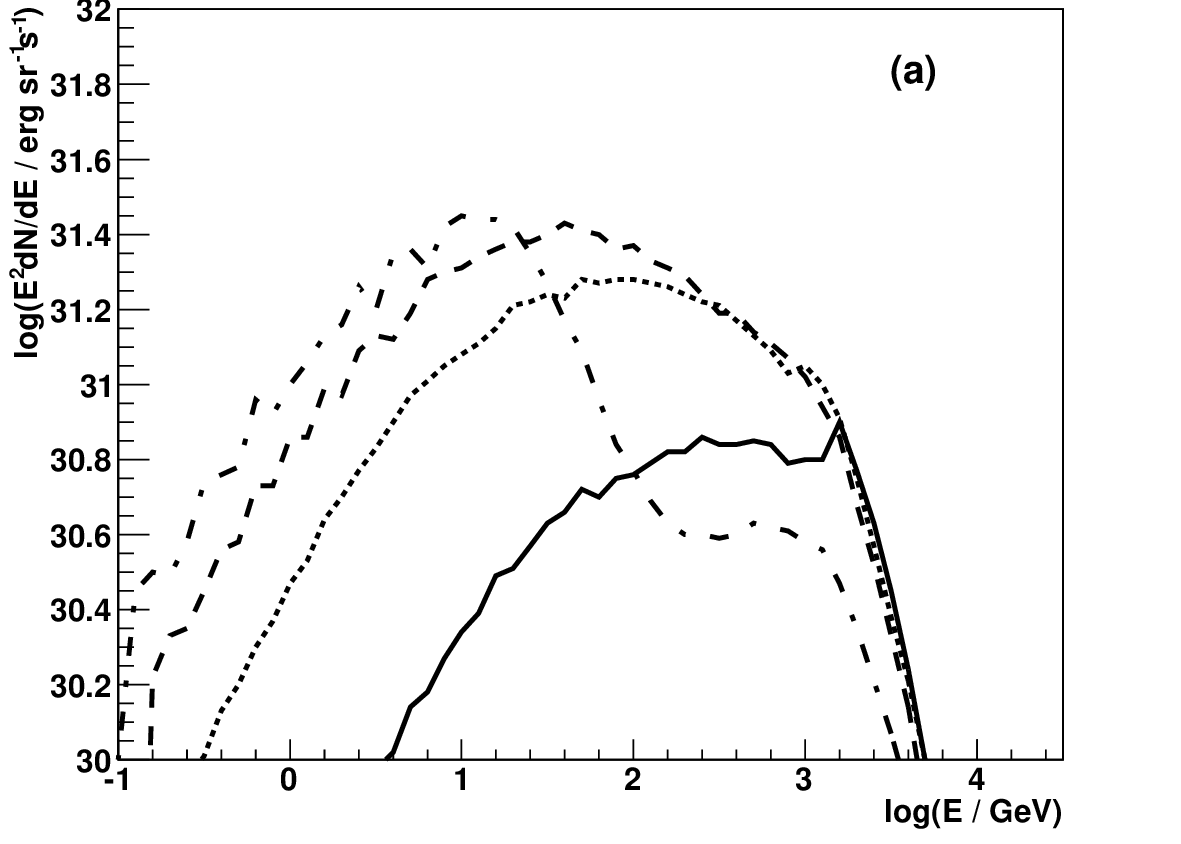}
\includegraphics{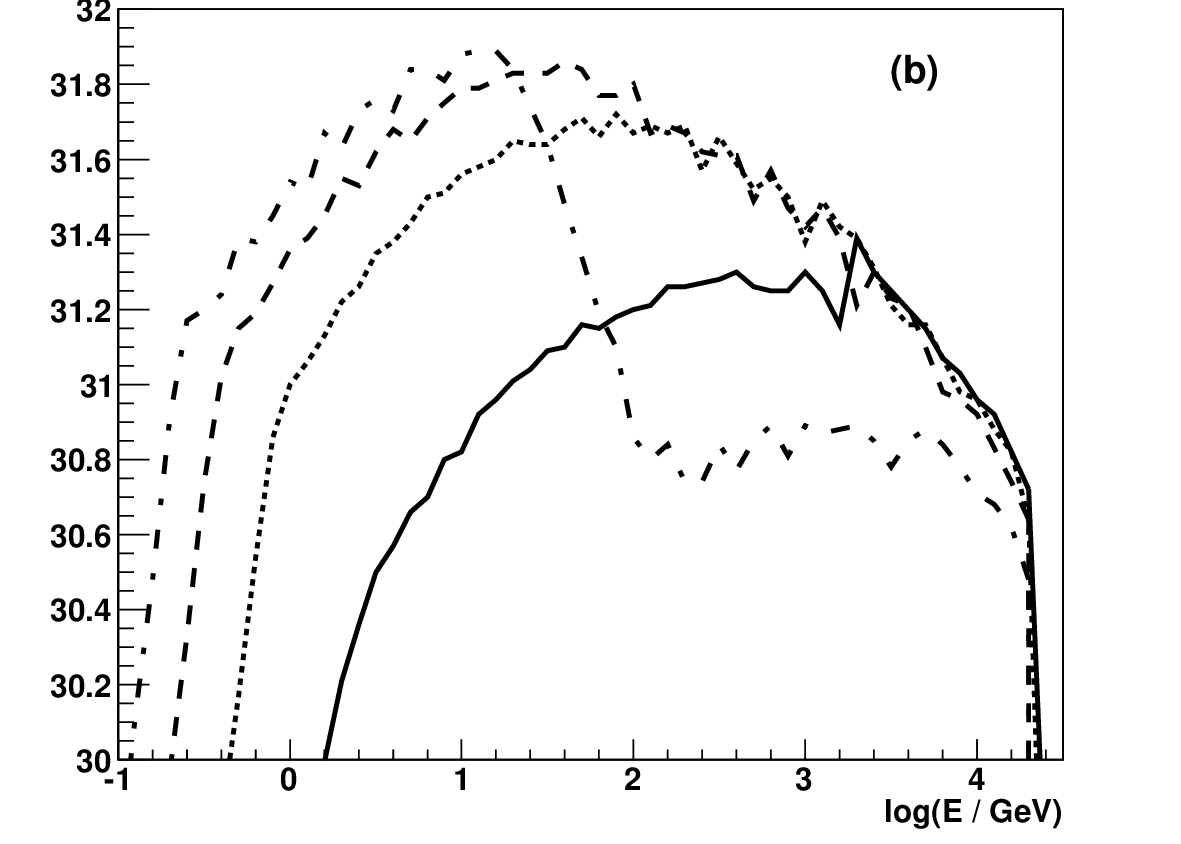}
\caption{The gamma-ray spectra, multiplied by the energy squared (Spectral Energy distribution - SED), emitted at specific directions on the sky, defined by the angle $\alpha$, from the vicinity of the massive star in the binary system containing pulsar PSR J2032+4127. Electrons are injected at the distance of the periastron passage of the pulsar, i.e. 
$D_{\rm peri} = 35R_\star$. The spectra, summed over the specific range of the cosine angle $\alpha$ which is measured from the outward direction defined by the stars, are shown for $-0.9 < \cos\alpha < -0.8$ (solid curve), $-0.3 < \cos\alpha < -0.2$ (dotted), $0.4 < \cos\alpha < 0.5$ (dashed), and $0.9 < \cos\alpha < 1.$ (dot-dashed). The parameters of the acceleration region are defined by the magnetization parameter of the pulsar wind, $\sigma = 0.1$, the velocity of the barion loaded wind is assumed to be  $v_{\rm mix} = 4.5\times 10^9$ cm s$^{-1}$, 
and the parameter $\eta = 0.1$. The acceleration coefficient of electrons is described by $\xi = (v_{\rm mix}/c)^2\approx 0.02$ (a) or $\xi = 0.1$ (b). Isotropic electrons are injected with the power law spectrum (the spectral index -2) up to the maximum energies, $E_{\rm max}$, which are determined by the balance of the acceleration time scale with the advection time scale or the synchrotron energy loss time scale. It is assumed that $10\%$ of the pulsar spin down power is transferred to the relativistic electrons.}
\label{fig4}
\end{figure}

We use mentioned above numerical code for the anisotropic IC $e^\pm$ pair cascade, initiated by relativistic electrons injected at some distance from the massive star, to calculate the synchrotron and the $\gamma$-ray spectra from the binary system containing PSR J2032+4127. The exemplary calculations of the $\gamma$-ray spectra observed from the binary system at specific directions are shown in Fig.~4.  
The two models considered in this figure differ in the description of the acceleration efficiency of the electrons in the collision region. In Fig.~4a, the acceleration coefficient is defined by the velocity of the mixed winds, $v_{\rm mix}$, according to $\xi = (v_{\rm mix}/c)^2$.  In Fig.~4b, $\xi$ is fixed to 0.1. The electrons can be accelerated with such a large efficiency already in the pulsar wind region when the velocity of the plasma flow is a significant part of the speed of light. The IC e$^\pm$ pair cascade $\gamma$-ray spectra  
strongly depend on the observation angle. When the direction towards the observer passes close to the massive star, the $\gamma$-ray emission in the GeV energy range clearly dominates over the TeV $\gamma$-ray emission. TeV emission is strongly suppressed due to the absorption effects. On the other hand, the $\gamma$-ray spectra at corresponding to the highest energies of the parent electrons only weakly depend on the observation angle since they are produced deep in the Klein-Nishina regime. Therefore, we conclude that the $\gamma$-ray emission should strongly depend on the geometry of the binary system and the location of the observer.

\begin{figure*}
%\centering
\vskip 9.truecm
\includegraphics{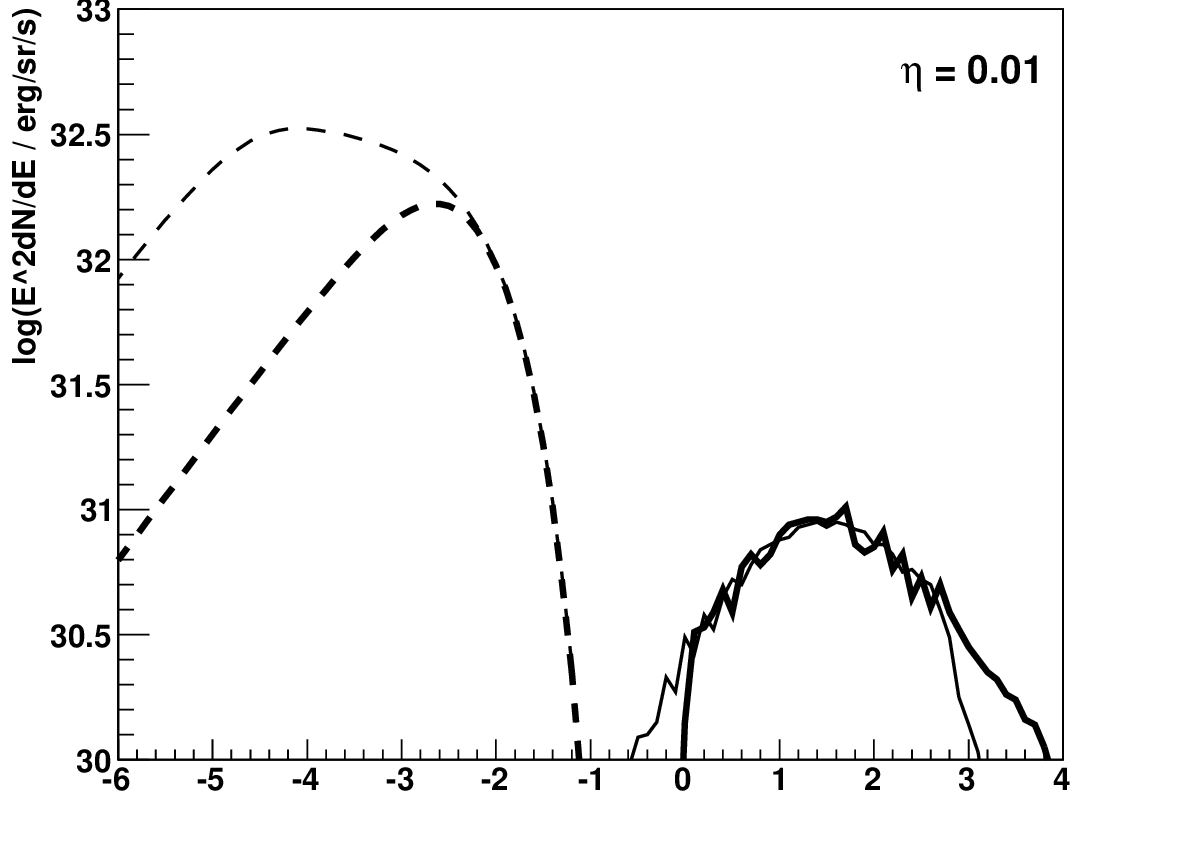}
\includegraphics{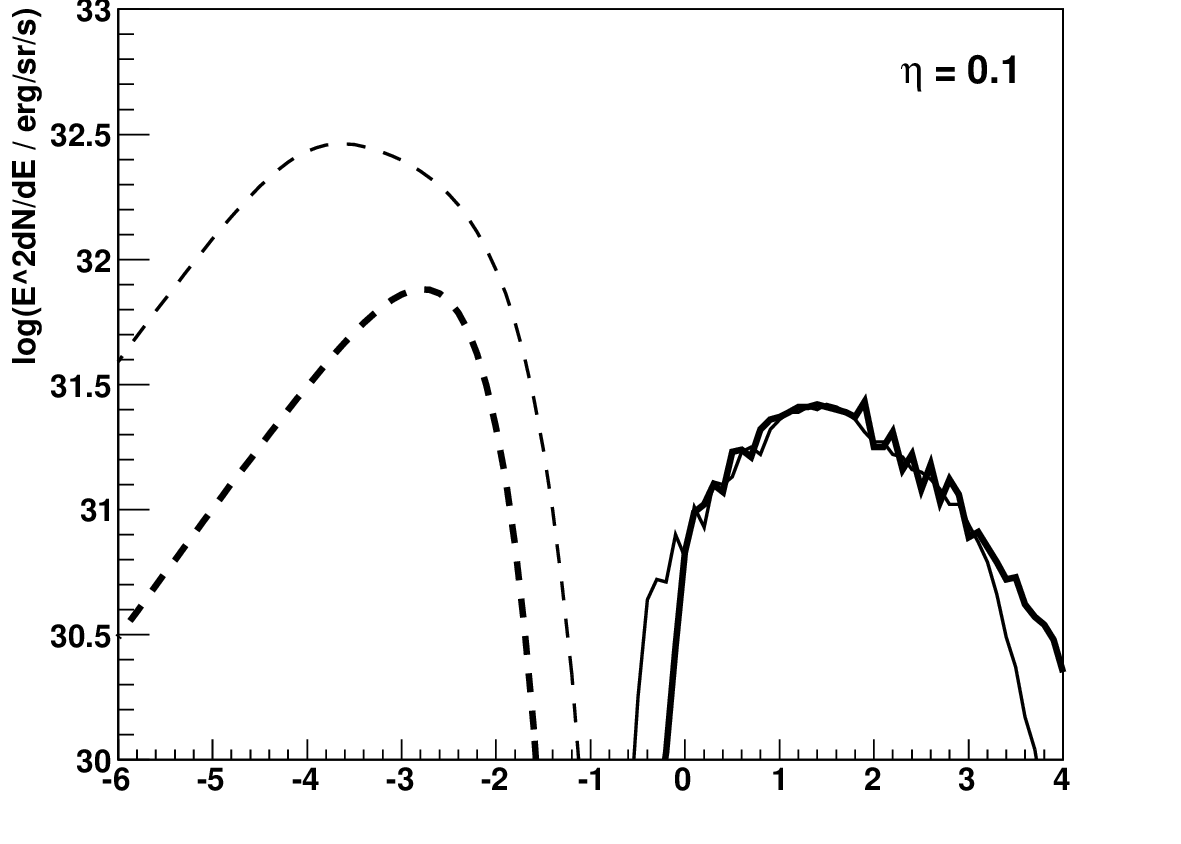}
\includegraphics{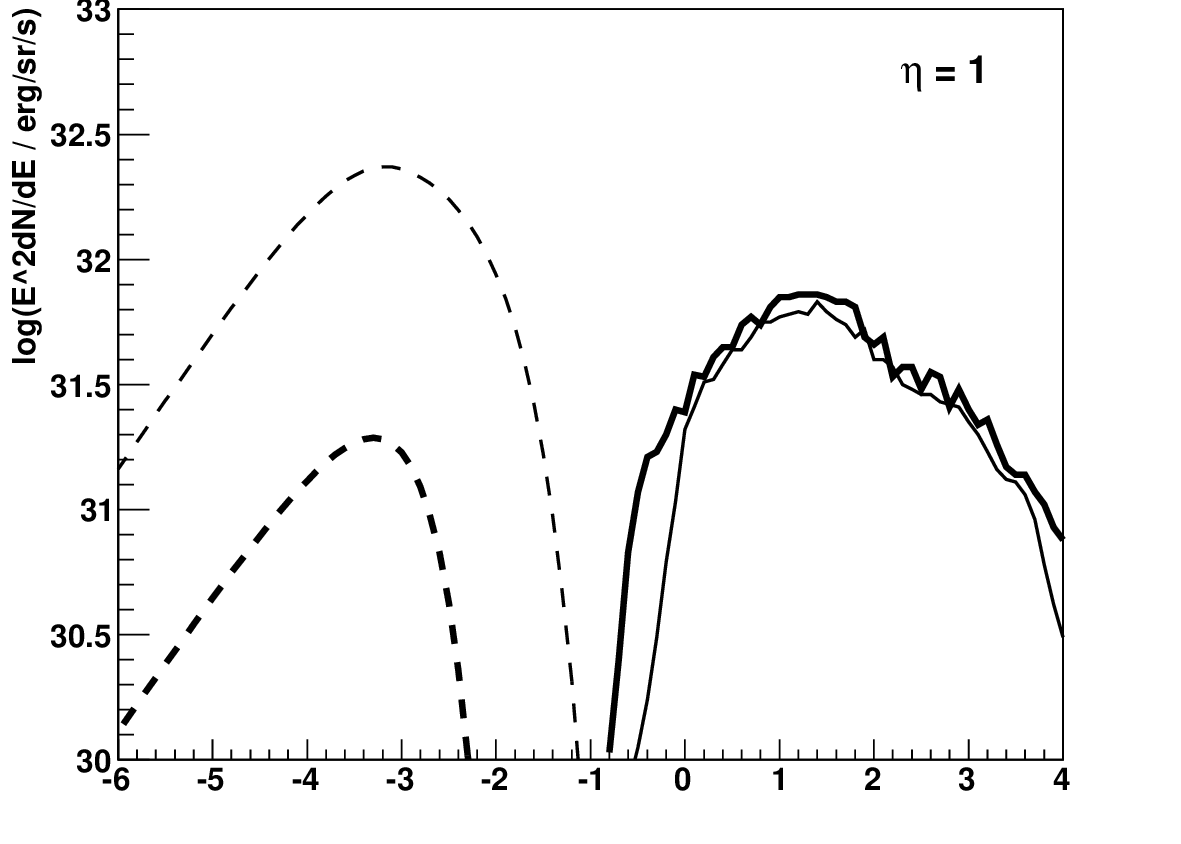}
\includegraphics{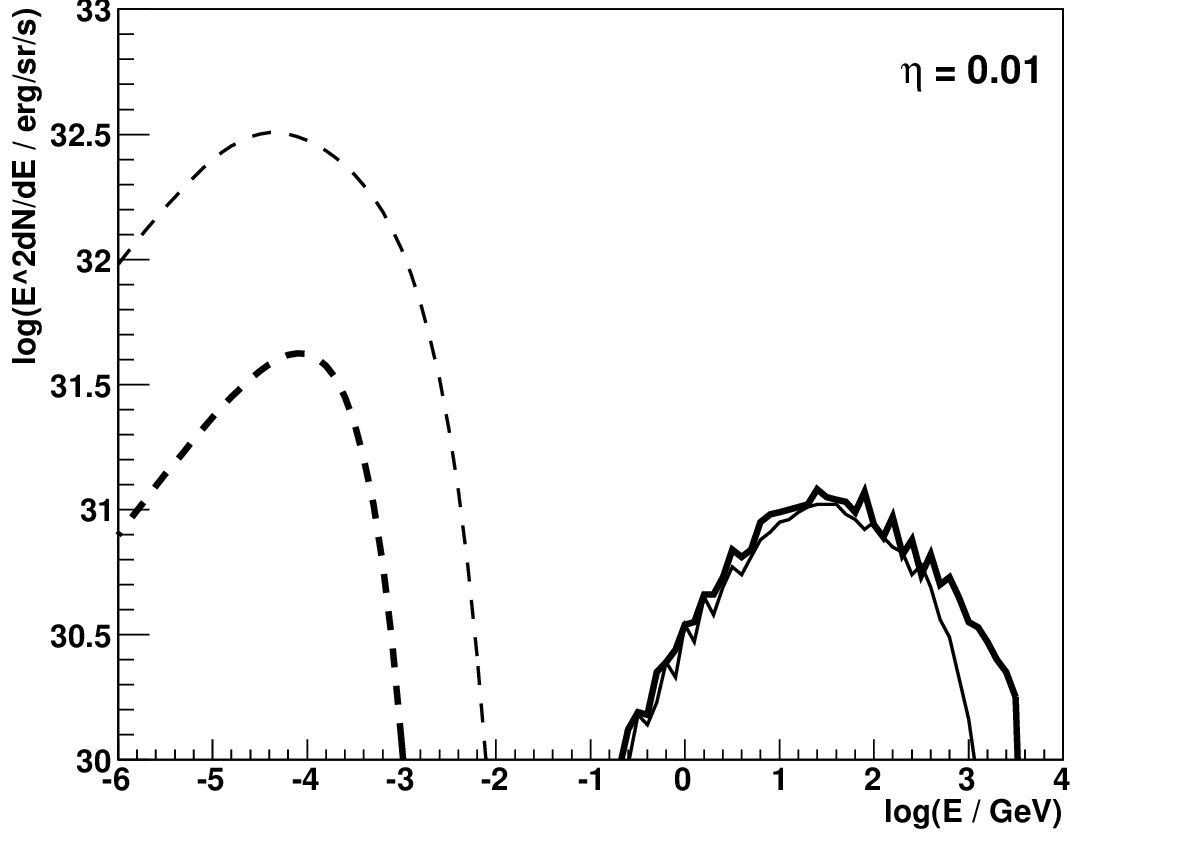}
\includegraphics{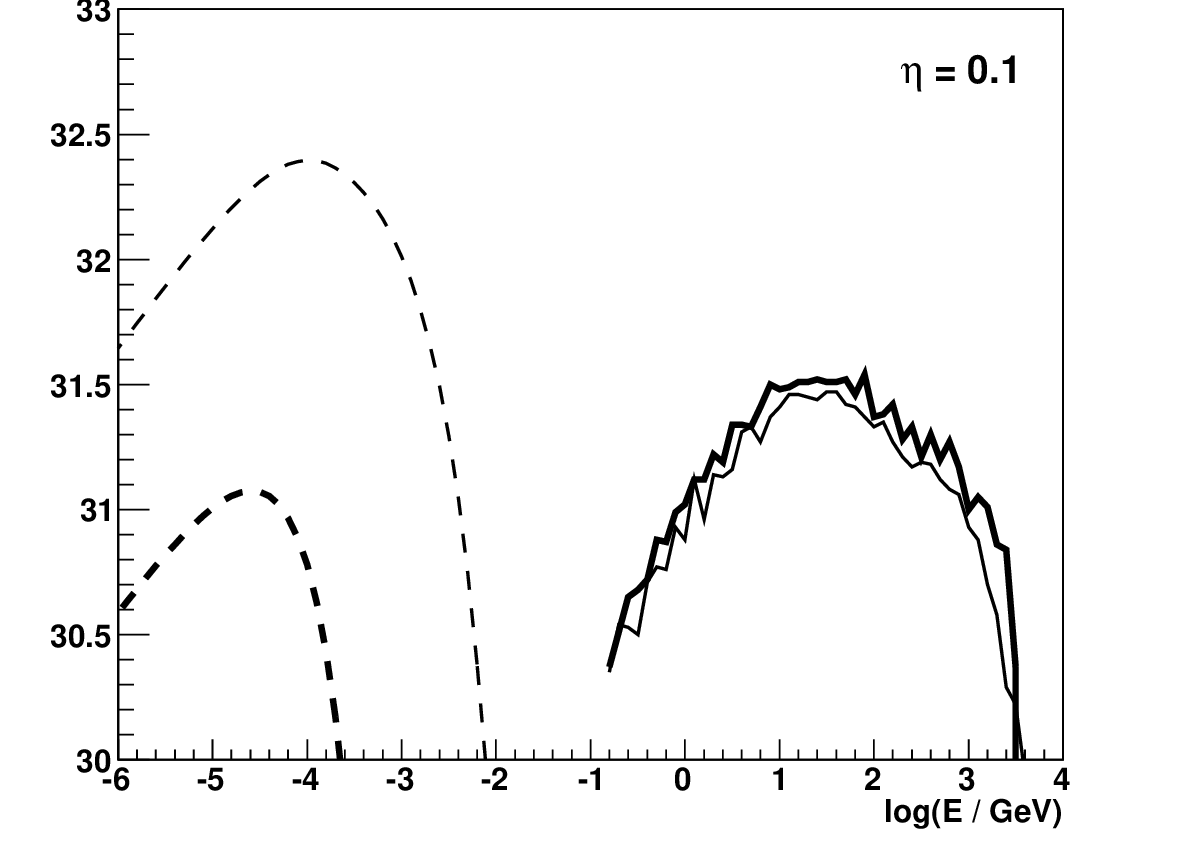}
\includegraphics{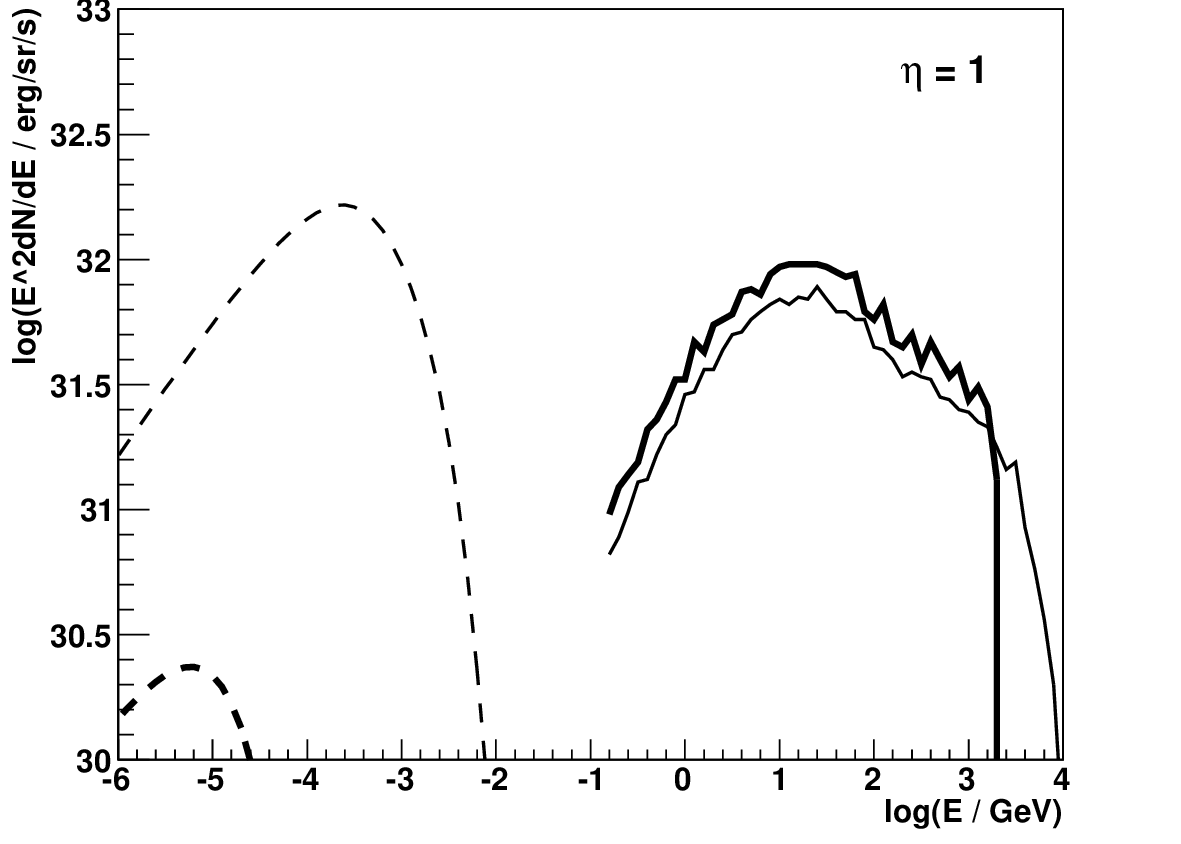}
\caption{SED of the synchrotron and the IC $\gamma$-ray spectra escaping towards the observer from the binary system containing pulsar PSR J2032+4127 at its periastron passage. Spectra are shown for different values of the parameter $\eta = 0.01$ (on the left), 0.1 (in the middle), and 1 (on the right). Two values of of the magnetization parameter of the pulsar wind are considered $\sigma = 0.003$ (thick curves), and $0.1$ (thin). The value of the acceleration coefficient of the electrons at the wind collision region is equal to $\xi = 0.1$ (upper figures) and 0.01 (bottom). The synchrotron spectra are marked by the dashed curves, and the $\gamma$-ray spectra from the IC process ($e^\pm$ pair cascade included) are shown for the inclination angle of the binary system $i = 60^\circ$ (solid curves). The other parameters describing the model, i.e. the binary system and the acceleration mechanism, are the same as in Fig.~4.}
\label{fig5}
\end{figure*}

In Fig.~5 we present more systematic investigations of the broad range spectra (synchrotron and IC) produced for 
different parameters describing the geometry of the binary system and the process of acceleration and propagation of the electrons.
We investigate the relative importance of the synchrotron and the IC $\gamma$-ray spectra as a function of the injection place of the relativistic electrons (described by parameter $\eta$). The spectra are shown for two magnetization parameters of the pulsar wind, 
$\sigma = 0.003$ (typical for the Crab-type pulsars, see e.g. \cite{kc84}) and $\sigma = 0.1$
(for the Vela-type pulsars $\sigma$ has been constrained in the range 0.05-0.5, see \cite{sd03}). The pulsar is located at the periastron passage and the inclination angle of the binary is fixed to $i = 60^\circ$. Note that at the periastron passage the change of the inclination angle in the range of $0^\circ - 90^\circ$ does not affect strongly the value of the angle $\alpha$ which determines the efficiency of the IC process (see Fig.~4). 
Note how strongly the synchrotron emission depends on the magnetization of the wind. The cut-off in the synchrotron spectra are determined by the physical process controlling the maximum energies of the accelerated electrons, i.e. either the synchrotron energy losses or the advection from the collision region. The intensities of the IC $\gamma$-ray spectra clearly increase with the value of $\eta$ since the radiation field from the companion star is stronger. However, the shape and intensity of the IC $\gamma$-ray spectra only weakly depend on the magnetization of the acceleration region since they are mainly determined by the advection time scale from the binary system.  

\begin{figure*}
%\centering
\vskip 4.5truecm
\includegraphics{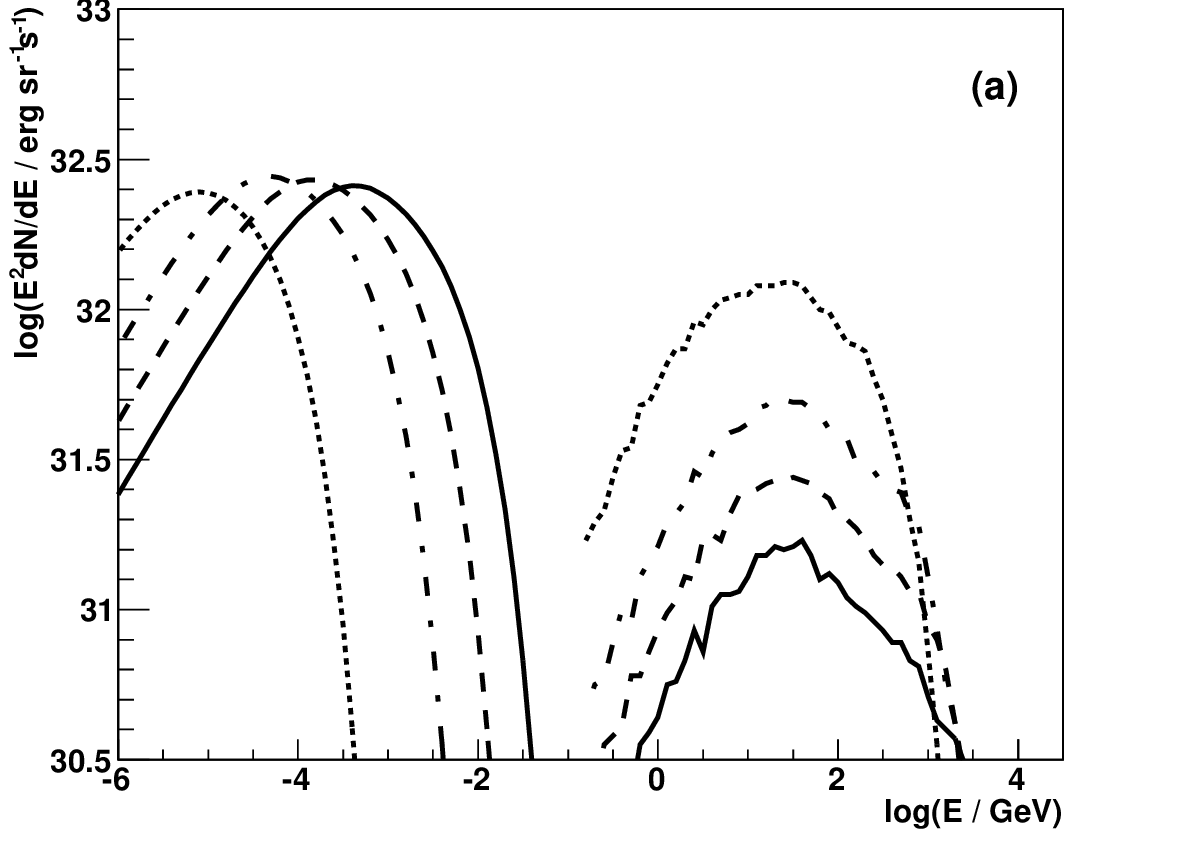}
\includegraphics{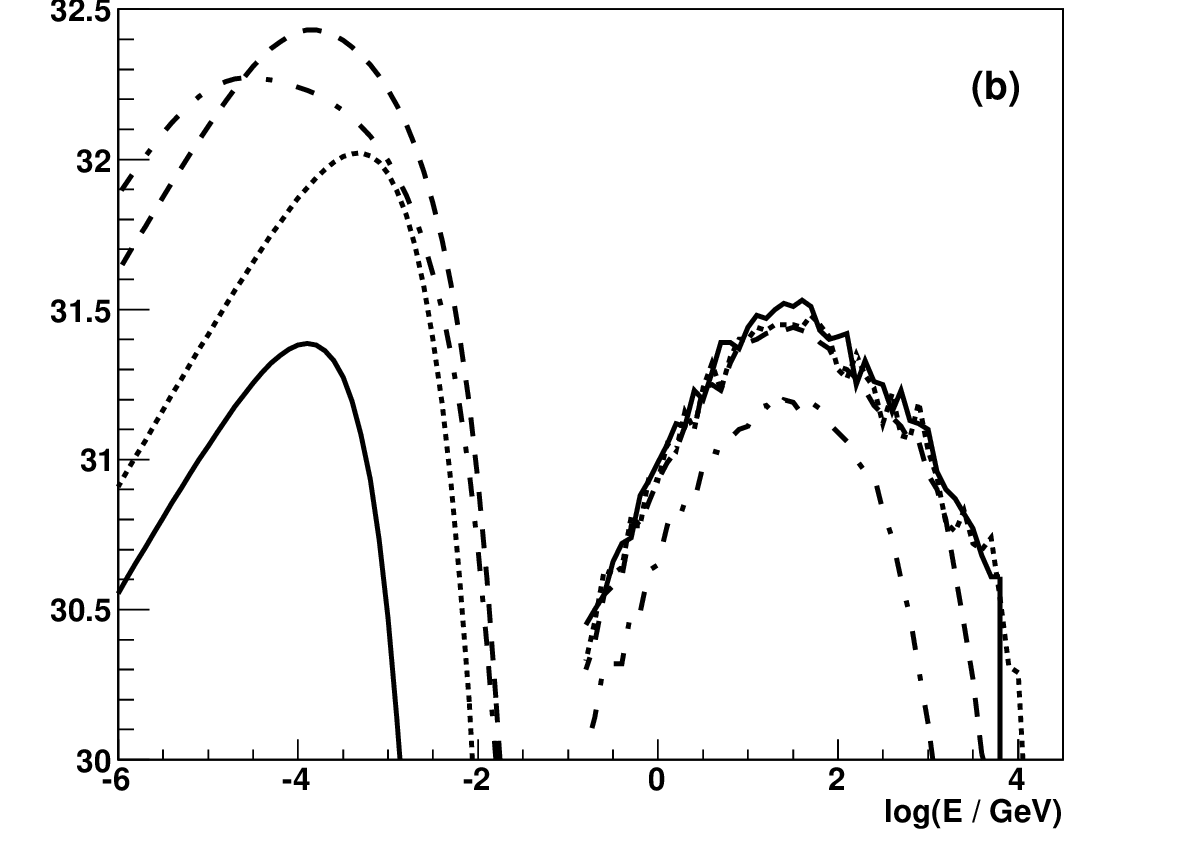}
\includegraphics{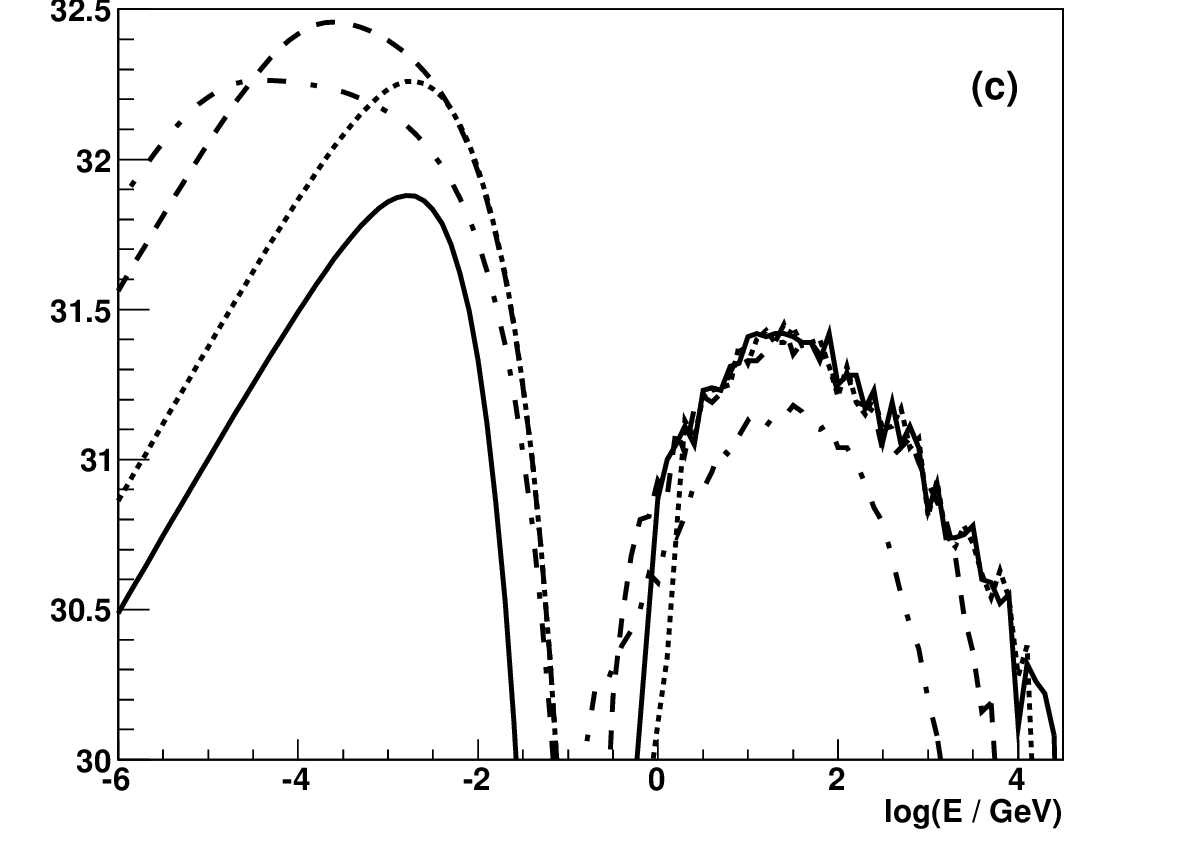}
\caption{As in Fig.~5 but for different values of the mixed wind velocity (figure a). The velocity of the pulsar wind, loaded with the matter of the stellar wind, has been fixed on $v_{\rm mix} = 10^{10}$ cm~s$^{-1}$ and 
$\xi = 0.1$ (solid curve), $4.5\times 10^9$ cm~s$^{-1}$ and $\xi = 0.02$ (dashed), $1.5\times 10^9$ cm~s$^{-1}$ and 
$\xi = 2.5\times 10^{-3}$ (dot-dashed), and $10^{8}$ cm~s$^{-1}$  and $\xi = 10^{-5}$ (dotted). The other parameters of the model are $\sigma = 0.1$, $\eta = 0.1$. The spectra for different magnetization parameters of the pulsar wind, for $\sigma = 0.003$ (solid curve), 0.01 (dotted), 0.1 (dashed), and 1 (dot-dashed), are shown in (b) (for $\xi = 0.02$) and in (c) (for $\xi = 0.1$). The other parameters of the calculations presented in figures (b) and (c) are fixed on $v_{\rm mix} = 4.5\times 10^9$ cm s$^{-1}$ and $\eta = 0.1$. The pulsar is located at the periastron passage and the inclination of the binary system is fixed on 
$60^\circ$. The parameters of the electron spectrum are as in Fig.~5.}
\label{fig6}
\end{figure*}

We also investigate the dependence of the non-thermal spectra, escaping from the binary system, on the velocity of the stellar wind for defined geometry of the collision region of the winds ($\eta = 0.1$) and the magnetization parameter of the pulsar wind ($\sigma = 0.1$), see Fig.~6. In these calculations the acceleration coefficient is related to the mixed wind velocity through the relation $\xi = (v_{\rm mix}/c)^2$. 
The structure of the wind around Be type stars is observed to be quite complicated. The equatorial wind is slow and dense and the polar wind is fast and rare. The detailed geometry of the wind, around the massive companion in PSR J2032+4127/MT91 213 binary system is, at present, unknown. It is also not clear in what type of the wind (polar, equatorial) the pulsar is immersed close to the periastron. Therefore, the wind velocities can change between a few tens km~s$^{-1}$ (equatorial) up to a few thousand km~s$^{-1}$ (polar). We show the dependence of the synchrotron and IC $\gamma$-ray spectra on the mixed wind velocities in Fig.~6a. The synchrotron spectra extend to lower energies for lower wind velocities since the maximum energies of the electrons depends on the mixed wind velocity through
the acceleration coefficient. On the other hand, the intensities of the IC $\gamma$-ray spectra are inversely proportional to the  values of the mixed wind velocity due to the dependence of the advection time of the electrons on the acceleration place. 
We also investigate the dependence of the spectra on the magnetization parameter of the pulsar wind for two models of the acceleration efficiency. In Fig.~6b, the acceleration efficiency depends on the mixed wind velocity. In Fig.~6c, it is fixed to $\xi = 0.1$. As expected, the synchrotron spectra dominate over the IC $\gamma$-ray spectra in the case of strongly magnetised pulsar winds. 
On the other hand, the intensities of the IC $\gamma$-ray spectra are almost independent for weakly magnetized winds. The IC $\gamma$-ray spectra decrease significantly and their high energy cut-off shifts to lower energies for strongly magnetized pulsar winds, i.e. $\sigma\sim1$. These spectral features are caused by the interplay between the different energy loss processes of the accelerated electrons (synchrotron and IC) and their escape from the radiation region due to the advection process.

\section{Comparison with high energy observations towards PSR J2032+4127}

The pulsar PSR J2032+4127 has been discovered in the direction of the unidentified HEGRA TeV $\gamma$-ray source
(TeV J2032+4130 \cite{aha02}). This TeV $\gamma$-ray source, with the extension of $\sim$6', has a flat spectrum (spectral index equal -1.9) and the flux of 5$\%$ of the Crab Nebula \cite{aha05b,alb08,ali14,lan04,kon07}. The TeV emission has been found to be persistent within the factor of $\sim$2 (between different experiments). The spread of measured fluxes might be due to the instrumental effects. More extended TeV $\gamma$-ray emission in the general direction of the Cygnus region has been also reported by the  MILAGRO and  the ARGO-YBJ experiments \cite{ab07,bar12}. 
The nature of the HEGRA TeV source remains unclear. Different scenarios suggest its relation to
a distant microquasar \cite{par07}, association with a cluster of massive stars observed in this direction \cite{but03,tor04,anh07,bed07} or even a distant blazar. However, recent discovery of the energetic pulsar in this direction, PSR 2032+4127, seems to support the hypothesis that TeV $\gamma$-ray source is caused by the pulsar wind nebula (PWN) formed by the unknown, at that time, young pulsar \cite{bed03,mur11}.

\begin{figure*}
%\centering
\vskip 8.5truecm
\includegraphics{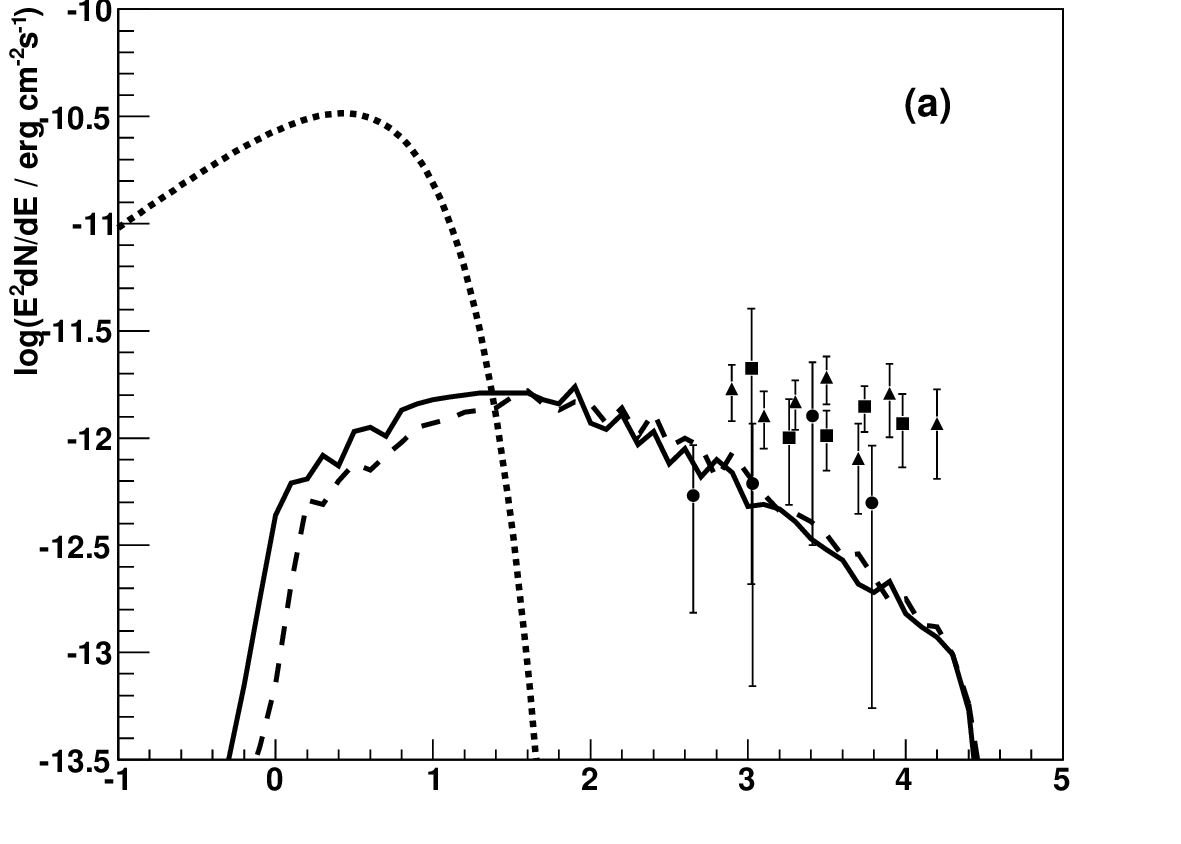}
\includegraphics{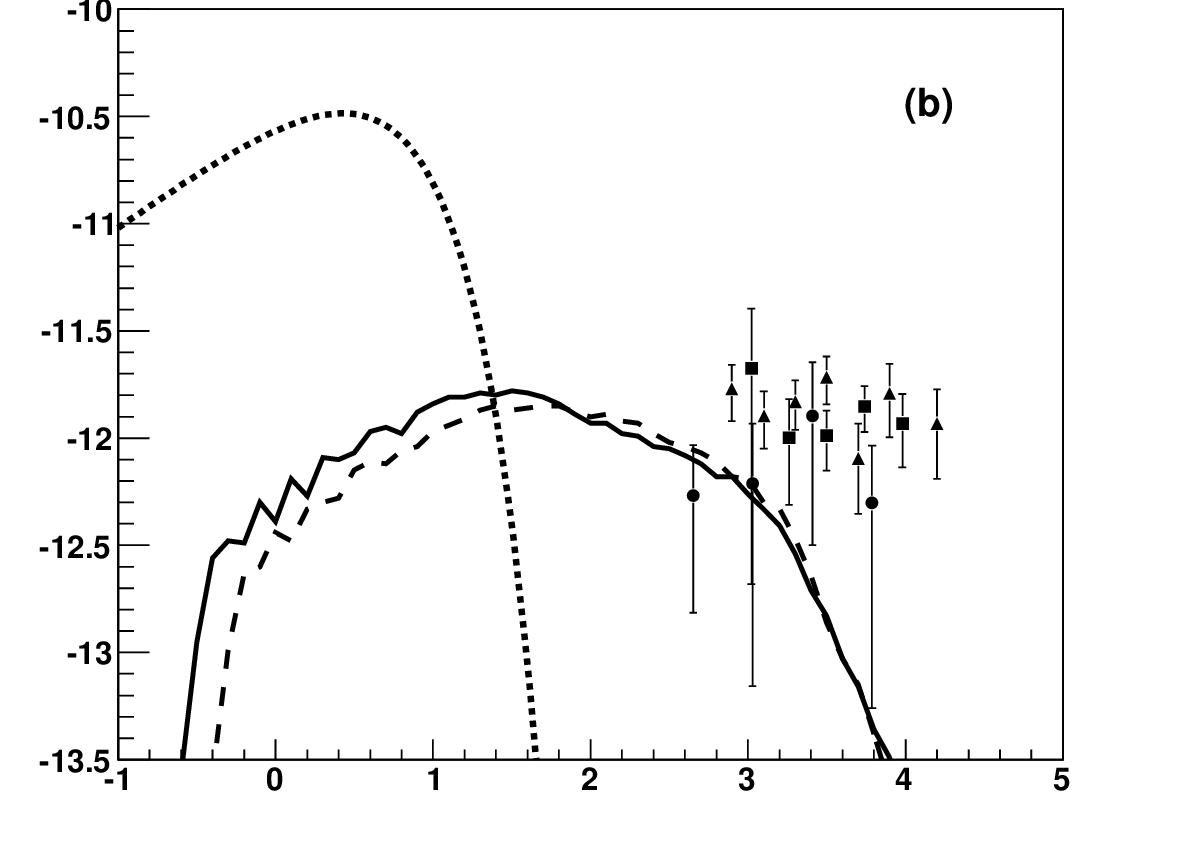}
\includegraphics{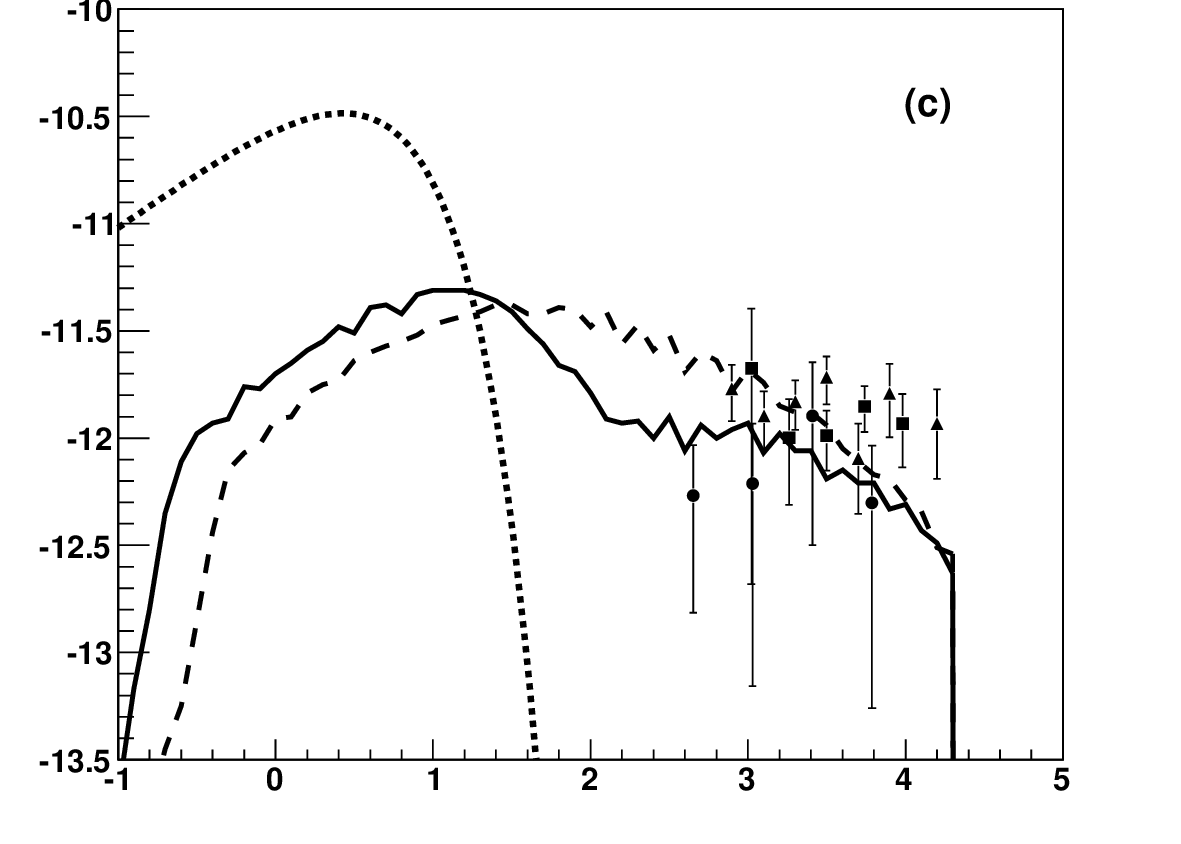}
\includegraphics{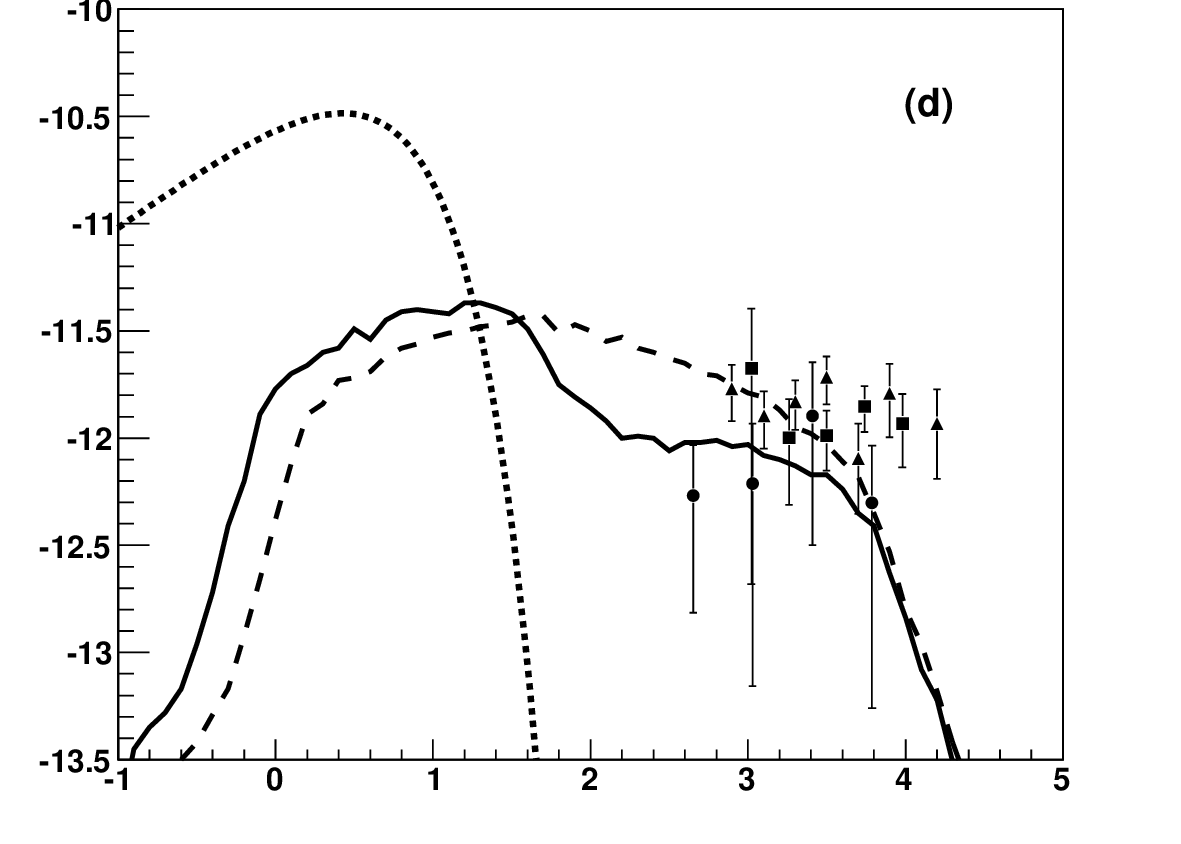}
\includegraphics{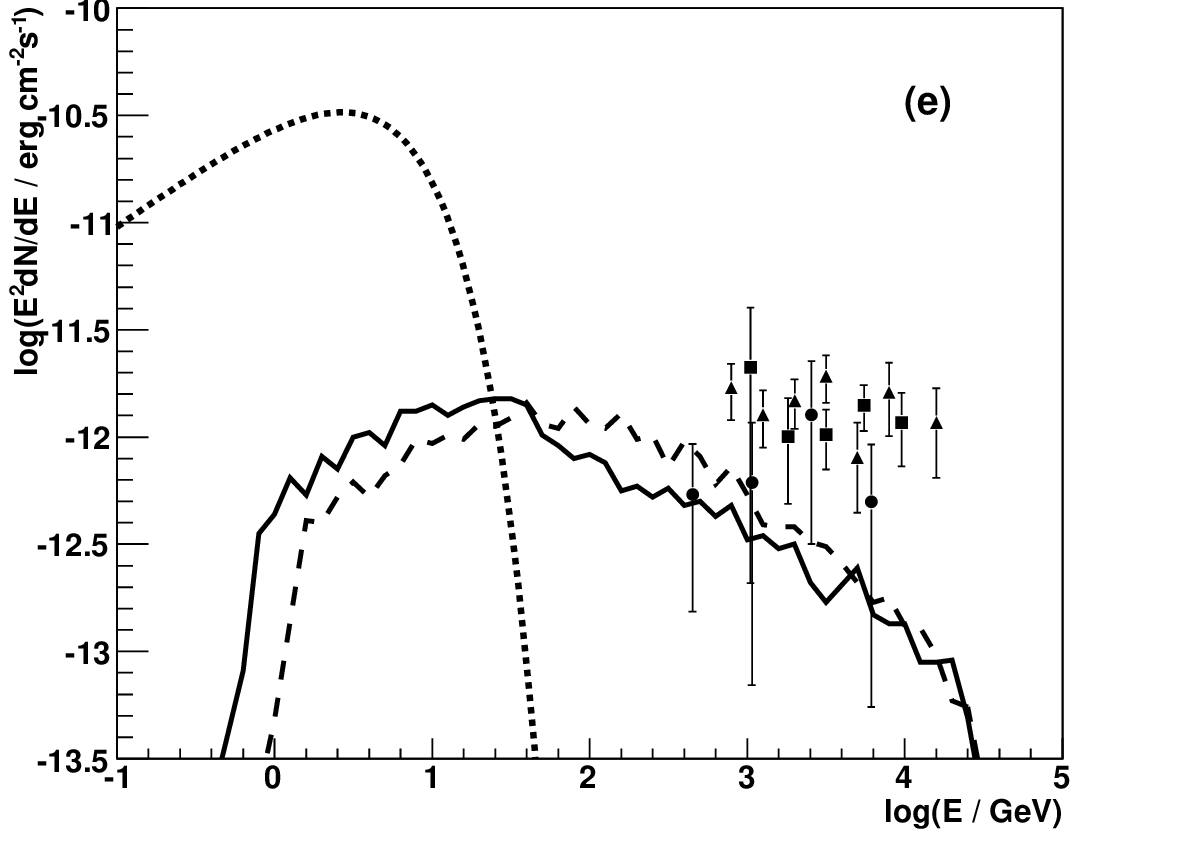}
\includegraphics{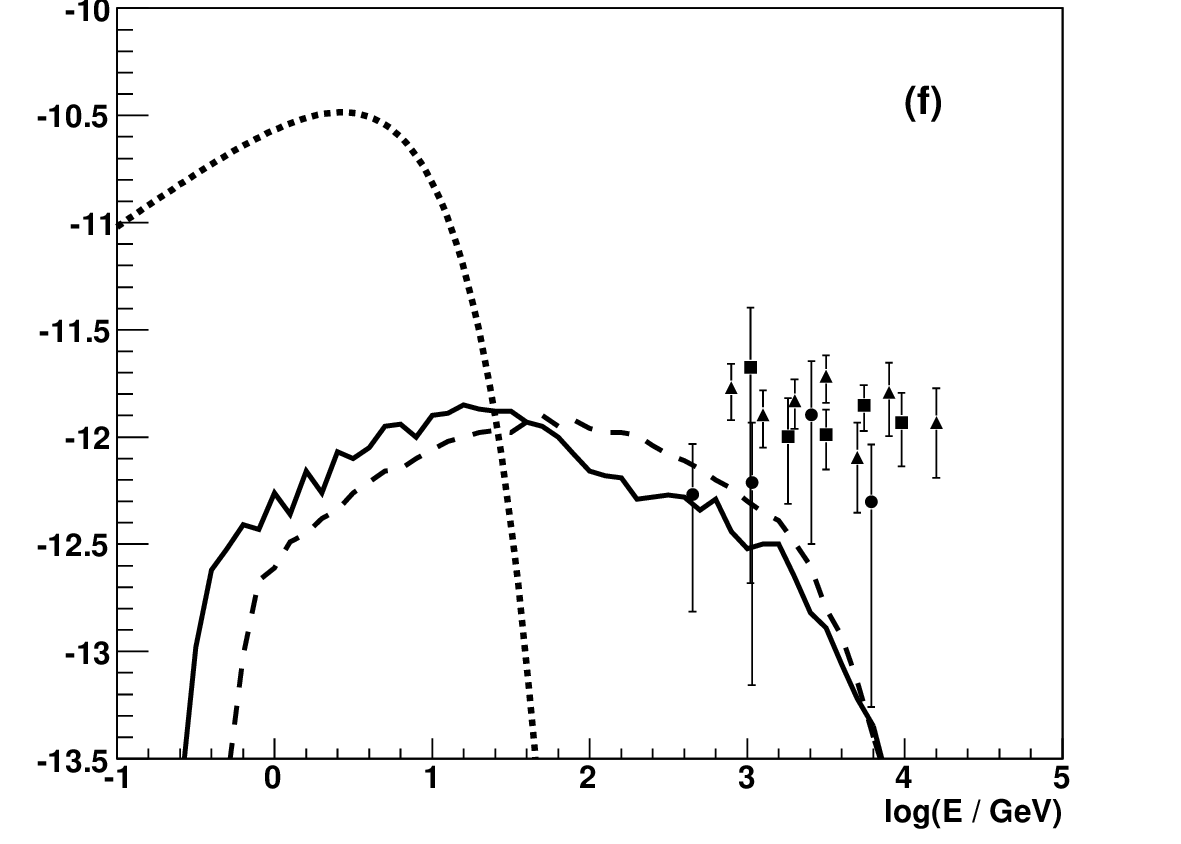}
\includegraphics{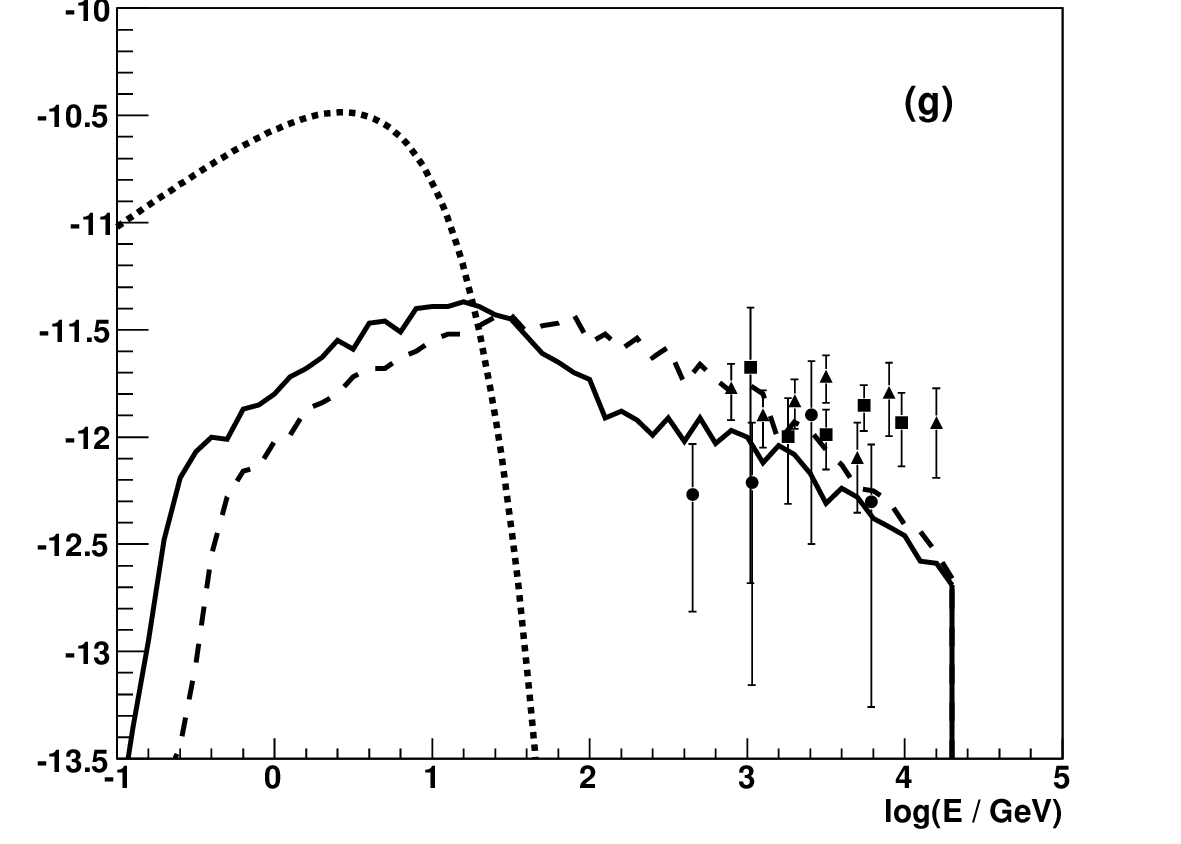}
\includegraphics{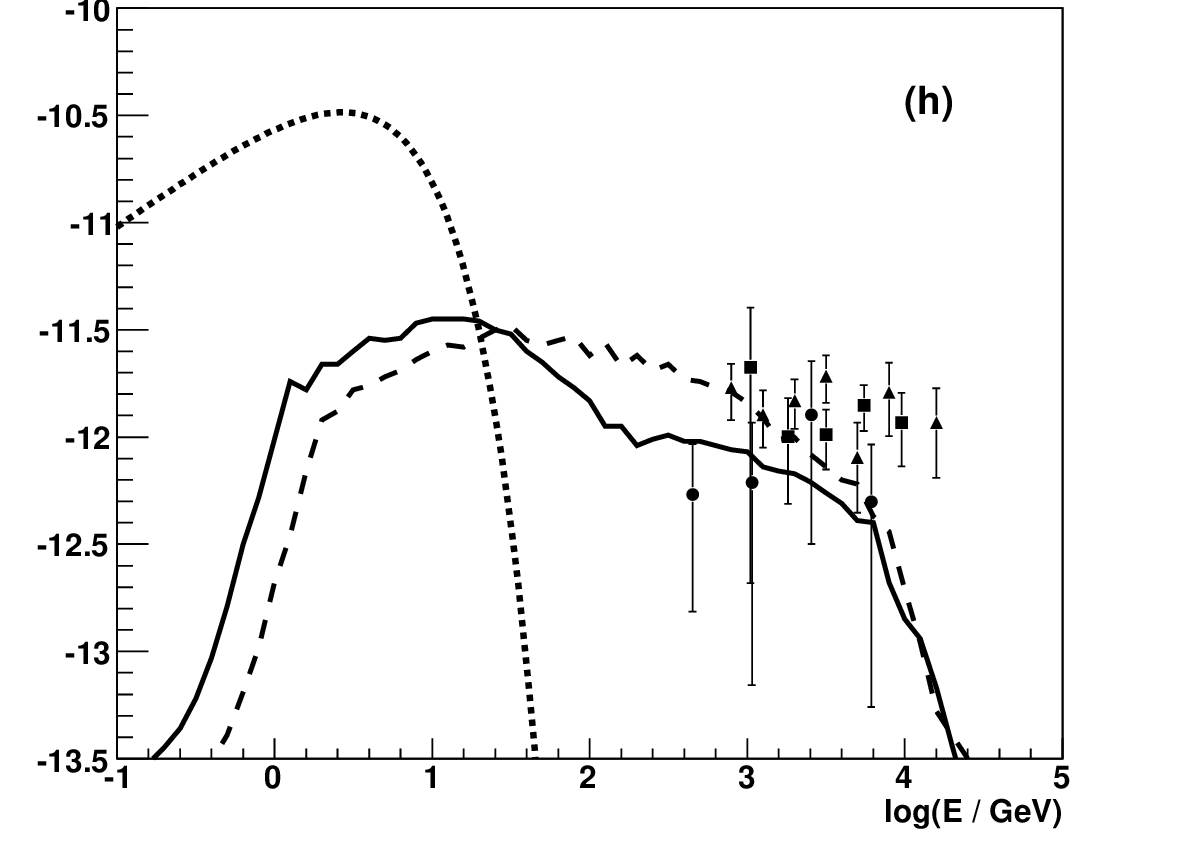}
\caption{The IC spectra (SED), at the periastron passage (upper figures) and at the superior conjunction of the pulsar (bottom figures) are compared with the TeV $\gamma$-ray spectrum of the extended source in the location of PSR J2032+4127, measured by HEGRA (black squares \cite{aha02}), MAGIC (circles \cite{alb08}), and VERITAS (triangles \cite{ali14}).
Figures (a) are for $\eta = 0.1$, $\xi = 0.1$, $v_{\rm mix} = 4.5\times 10^9$ cm~s$^{-1}$ and $\sigma = 0.003$; (b) for $\eta = 0.1$, $\xi = 0.1$, $v_{\rm mix} = 4.5\times 10^9$ cm~s$^{-1}$ and $\sigma = 0.1$; (c) for $\eta = 1$, $\xi = 0.1$, $v_{\rm mix} = 
7\times 10^9$ cm~s$^{-1}$ and $\sigma = 0.003$; and  (d) for $\eta = 1$, $\xi = 0.1$, $v_{\rm mix} = 7\times 10^9$ cm~s$^{-1}$ and $\sigma = 0.1$. The parameters of the electron spectra are as in previous figures.  The spectra are calculated for the inclination of the binary system equal to $i = 60^\circ$ (solid curves) and $i = 0^\circ$ (dashed curves). The pulsed $\gamma$-ray spectrum of PSR J2032+4127 \cite{ho17} is marked by the dotted curve.}
\label{fig7}
\end{figure*}

The orbit of the pulsar PSR J2032+4127 is so eccentric that for most of the time the radiation of the companion star is not able to provide important target for the relativistic electrons accelerated in the PWN. However, the IC scattering of the stellar radiation might become important when the pulsar is close to its periastron (as observed in the case of PSR1259-63/LS2883). In order to determine the contribution of the emission from the binary system to the already observed TeV $\gamma$-ray emission from the nebula, we confront the TeV $\gamma$-ray spectrum of the HEGRA source, i.e. measured when the pulsar is far from the periastron, with the TeV $\gamma$-ray emission due to the comptonization of the stellar radiation by the electrons accelerated at the collision region of the pulsar and the stellar winds. In Fig.~7, we show the $\gamma$-ray spectra calculated for the pulsar at the periastron (the strongest radiation field of the companion star) and at the superior conjunction (direction between electrons and the observer with the closest approach to the stellar surface), for specific range of parameters describing the acceleration process and the interaction of the electrons with the stellar radiation. The range of phases between these two locations of the pulsar on the orbit provides the best conditions for the production of $\gamma$-rays in the direction of the observer on the Earth.
In these exemplary calculations, we assume that the energy transferred from the pulsar to the relativistic electrons is equal to $10\%$. This value is close to the upper limit if the electrons are accelerated in the Fermi process.
The spectra are calculated for the wide range of the inclination angles of the binary system ($i = 0^\circ$ and $60^\circ$) since the value of the inclination angle is at present unknown. In fact, the calculated 
$\gamma$-ray spectra only weakly depend on the considered range of the inclination angles of the binary system (see dashed and solid curves in Fig.~7). The contribution to the TeV $\gamma$-ray emission, produced within the binary system at the periastron, to the extended TeV emission from the nebula (the HEGRA TeV $\gamma$-ray source), becomes important when the parameter $\eta$, describing the distance of the collision region of the winds from the companion star, is not very far from unity (see Figs.~7c,d,g,h). Large value of $\eta$ is expected when
the pulsar, close to its periastron, is immersed in the region of the polar wind of the companion Be type star. The polar wind can reach the velocity of the order of $10^3$ km~s$^{-1}$. However, the mass loss rate of the polar wind can be about 2-3 orders of magnitude lower than that in the equatorial wind. In such a case, 
$\eta$ can reach values close to unity. The collision region can also appear close to the star if the flow 
of the winds is determined by the presence of the dense decretion disk. Such situation has been analysed in the analytical way in \cite{sb08} and through the detailed hydrodynamic simulations in \cite{ok11,ta12}. It is expected in the case of the binary system PSR B1259-63/LS2883. This binary contains a well documented Be type star, in which the geometry of a decretion disk is well defined. 

Our present calculations show that the emission from the binary system PSR J2032+4127/MT91 213 at the periastron passage can dominate over the extended emission from the HEGRA source, but only at energies lower than a few TeV (see Fig.~7c and 7d). 
In order to investigate whether the IC emission at the periastron passage can be detected on top of the already known extended source at this position we have used MAGIC performance reported in Aleksic et al. \cite{ale16} implemented in \emph{MAGIC source simulator} tool \footnote{$https://magic.mpp.mpg.de/fileadmin/user\_upload/mss.cpp$}. We have assumed that the known source has an extension of $\sim0.08^\circ$ and the IC emission at the periastron passage will be point-like. Due to the much improved sensitivity of MAGIC stereo system with respect to the mono observations reported in \cite{alb08}, about 26\,h of observations with the current MAGIC stereo system would be needed to re-detect the extended source and measure its spectrum with a similar accuracy. Next, using the same procedure, we simulate the IC spectra presented in Fig.~7. We increase the uncertainties of the IC spectrum corresponding to adding and subtracting the background source. We find that for the case of Fig.~7c and~7d (inclination angle $i = 0^\circ$) about 10\,h of observations during periastron passage should be sufficient to significantly disentangle the IC emission component in the energy range $\sim 200-1000$\,GeV. The IC component at the energies of $\sim 500$\,GeV is three times more pronounced than the detected by the MAGIC telescopes extended source. Therefore, the subtraction of the two observations will not be burdened with a large systematic uncertainty.

Electrons with multi-TeV energies are not able to scatter efficiently stellar radiation due to the Klein-Nishina effects and also also due to their dominant synchrotron energy losses. The escape process of the electrons from the acceleration region and the synchrotron energy losses limit their maximum energies just above $\sim$10 TeV. Therefore, the TeV 
$\gamma$-ray spectra from the binary system are not expected to extend through the multi-TeV energy range. The features of the TeV 
$\gamma$-ray emission, for the pulsar located at the superior conjunction, are quite similar to those observed at the periastron (but on a slightly lower level). Both these locations are separated by a few weeks.
Therefore, the TeV $\gamma$-ray emission from the binary system between periastron and superior conjunction is expected to be on the level observed from the extended HEGRA TeV $\gamma$-ray source (see Fig~5 for the parameters of the model when such situation is possible). We do not consider here the details of the TeV $\gamma$-ray light curve from this binary system since outside 
the above mentioned period the $\gamma$-ray emission is expected to be on a clearly lower level. Outside the region between the periastron and the superior conjunction, the pulsar (and relativistic electrons) stays at a clearly larger distance from the companion star and/or they are located more in the front of the companion star. Such a geometry is not favoured for the efficient production of $\gamma$-rays by scattering the stellar radiation. 

The TeV $\gamma$-ray emission (shown in Fig.~7) has been calculated for the case of a relatively large velocity of the pulsar wind, loaded with the matter of the stellar wind. Such relatively fast mixed winds are expected when 
the pulsar is emerged in the fast and rare polar wind of the Be star. If the mixed pulsar wind velocity is an order of magnitude lower, but the parameter $\eta$ is still close to unity, then the emission levels should be enhanced by a factor of $\sim$3 (see Fig.~6a). On the other hand, the level of the TeV $\gamma$-ray emission is only weakly dependent on the magnetization parameter of the wind, $\sigma$, provided that its value is clearly lower than unity. In contrast, to the emission of the synchrotron radiation strongly depends on $\sigma$ (see Figs.~6b,c).

\section{Discussion and conclusion}

We have presented the results of the calculations of the $\gamma$-ray emission from the anisotropic IC $e^\pm$ pair cascades initiated by relativistic electrons in the vicinity of the massive star within the binary system containing PSR J2032+4127. 
We show that the $\gamma$-ray emission from the binary system can overcome the level of the TeV emission from the HEGRA TeV $\gamma$-ray source, provided that the electrons are relatively slowly advected from the vicinity of the massive star. Such a situation is expected when the stellar and pulsar winds mix effectively in the wind collision region and the velocity of the stellar wind is not very large 
(i.e. $v_\star\le 10^8$ cm s$^{-1}$). Moreover, 
the collision region has to be close to the massive star. It happens when the pressure of the stellar wind does not dominate completely over the pressure of the pulsar wind, i.e. the parameter $\eta$ describing the wind collision region is not far from unity. This condition is satisfied when the mass loss rate of the massive star is $\dot{M}_\star\le L_{\rm PSR}/(\eta cv_\star)\approx 10^{-9}/(\eta v_8)$~M$_\odot$~yr$^{-1}$ (based on Eq.~1). 
For $\eta\sim1$, the collision region of the winds is far from the pulsar and relatively close to the stellar surface. Then, the magnetic field in the mixed pulsar wind is relatively weak (due to a larger distance from the pulsar). Therefore, the energy losses of the relativistic electrons on the synchrotron process are reduced and energy losses on the IC process are enhanced. It is expected that
the IC $\gamma$-ray fluxes shown in Figs.~7 are clearly larger if the mixed wind velocities are below the value 
considered in this figure (see dependence on the mixed wind velocity in Fig.~6a). However, the efficiency of the electron acceleration can be lower than the assumed $10\%$ since only a part of the wind region can provide energy for the acceleration process. Therefore, the comparison of our calculations with the future 
observations of the binary system PSR J2032+4127/MT91 213 during periastron can allow to reduce the range of allowed values of the parameters determining the geometry of the collision region $\eta$, the efficiency of the electron acceleration and the velocity of the mixed winds. Note that these parameters can be also constrained by independent observations in other energy ranges. 
Such constraints will allow determination of the acceleration efficiency of electrons within this binary system 
based on the TeV $\gamma$-ray observations and modelling presented in this paper. 

\begin{figure}
%\centering
\vskip 6.8truecm
\includegraphics{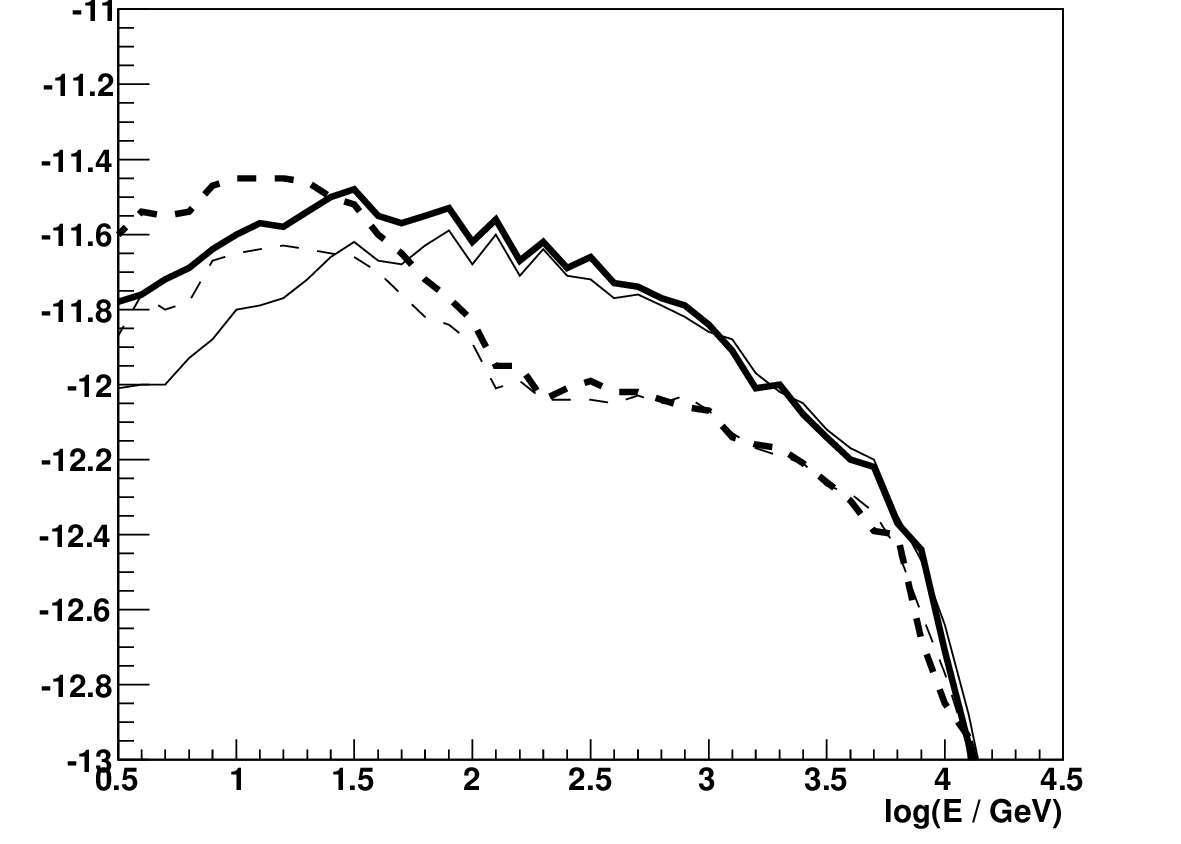}
\caption{The comparison of the $\gamma$-ray spectra from the anisotropic IC $e^\pm$ pair cascade developing within the whole volume of the binary system (thick curves) and the IC $\gamma$-ray spectrum produced (directly towards the observer without cascading) in the first interaction of primary electrons (thin curves) Two inclination angles are shown $i = 0^\circ$ (solid curves) and $60^\circ$ (dashed). The other parameters of the model are the following $\eta = 1$, $\xi = 0.1$, $v_{\rm mix} = 7\times 10^9$ cm~s$^{-1}$ and $\sigma = 0.1$ and the parameters of the electron spectrum are as in Fig.~7.}
\label{fig8}
\end{figure}

The $\gamma$-ray spectra shown in Fig.~7 are calculated in terms of the specific anisotropic IC $e^\pm$ pair cascade model. We compare the $\gamma$-ray spectra produced in the cascade process with the $\gamma$-ray spectra produced by the primary electrons (without any cascading effects) directly to the observer in Fig.~8. 
It is clear that secondary $e^\pm$ pairs contributes significantly to the energy range below 
$\sim$ 100 GeV. This feature is qualitatively similar to that already noted in the case of the 
$\gamma$-ray production in the TeV $\gamma$-ray binary system LS~I~+61$^\circ$~303 
\cite{bed06}.  We conclude that the $\gamma$-ray emission from the secondary $e^\pm$ pairs produced in the cascade has an important effect on the relative $\gamma$-ray luminosities expected in the GeV and TeV energy range.

Observations of the X-ray emission from the binary system  PSR J2032+4127/MT91 213 was showing a clear increase of the flux in the Swift/XRT energy range (between 0.3-10 keV) during the year 2016, reaching the value of $\sim$2.5$\times 10^{-12}$ erg cm$^{-2}$s$^{-1}$\cite{tak17}. Such trend is generally consistent with the predictions of the considered here model. 
The TeV gamma-ray flux predicted by our model at the periastron passage is on the level of 
$\sim$(1-3)$\times 10^{-12}$ erg cm$^{-2}$s$^{-1}$ (see Fig.~7). On the other hand, the expected X-ray flux can be even an order of magnitude larger 
(see Figs.~5 and 6). So then, we conclude that our model predicts still increasing X-ray flux when the pulsar
approaches the periastron. It can reach the value of $\sim$(1-3)$\times 10^{-11}$ erg cm$^{-2}$s$^{-1}$ at the periastron.

Our calculations show that the conditions for the $\gamma$-ray production within the binary system PSR J2032+4127 close to the periastron are less favourable than those in the case of the binary system PSR B1259-63/LS2883. This is generally expected since the energy loss rate of the pulsar PSR J2032+4127 is a factor of about 3 lower than the pulsar PSR B1259-63. Moreover, during the periastron passage, PSR B1259-63 is at the distance of 23$R_\star$ from the companion star and  PSR J2032+4127 at the distance of 35$R_\star$. Closer location of the pulsar PSR B1259-63 to the companion star at the periastron and its larger energy loss rate shifts the wind collision region, and thus also the place of electron acceleration, closer to the companion star. The companion star of PSR B1259-63 is also more luminous ($L_\star\approx 2.5\times 10^{38}$ 
erg~s$^{-1}$\cite{neg11}), than the companion star of PSR J2032+4127 ($L_\star\approx 5.8\times 10^{37}$ erg~s$^{-1}$). Therefore, the stellar radiation field in the region of electron acceleration is expected to be stronger in the case of this system. Moreover, the power transferred from the pulsar to relativistic electrons can be also larger due to the larger energy loss rate of PSR B1259-63. Therefore, it is not surprising that the TeV $\gamma$-ray emission from the binary system of PSR J2032+4127 is predicted to overcome the steady emission from the large scale nebula around the pulsar only at the special conditions discussed above. Detailed observations of the binary system PSR J2032+4127/MT91 213 by the Cherenkov telescopes close to the periastron, linked with
the future multi-wavelength studies of the stellar wind from the companion star, should provide interesting independent constraints on the proprieties of the pulsar winds, the conditions for the acceleration of particles, and the details of their radiation mechanisms.
At present such detailed comparison of both binary systems is not possible due to the lack of the knowledge on
the possible contribution of additional radiation fields and their geometry (e.g. a stellar disk) in the binary PSR J2032+4127/MT 91 213. 
Note that the present theoretical models, dedicated to the binary PSR B1259-63/LS2883, are not able to explain the details of the complicated features of its $\gamma$-ray light curve, such as the late GeV $\gamma$-ray peak or its variability in the TeV $\gamma$-rays between different revolutions. In fact, it is not so surprising since the content of the binary (the wind parameters and the magnetic field structures) and also the geometry, which plays the crucial role in location of the emission region, are not fully known.

Results of the TeV $\gamma$-ray spectra from the anisotropic IC $e^\pm$ pair cascades presented above can be affected if other soft radiation fields contribute significantly to the dominant radiation produced by the surface of the companion star. In fact, in the case of the Be type stars an additional lower energy radiation component can be produced in the decretion disk. In the case of the binary system PSR J2032+4127/MT91 213 the possible role of the decretion disk radiation has been discussed in an approximate way in \cite{tak17}. 
The contribution of this radiation field cannot be taken into account reliably in our present calculations since the detailed geometry of the disk and its parameters are not yet determined in the case of the companion star of the pulsar PSR J2032+4127. Note that the cascade model considered by us strongly depends on the geometry of the soft radiation field, which is scattered by relativistic electrons. 
Present observations of the equatorial disk allow us to constrain its outer radius only below $\sim (3-7)\times 10^{12}$ cm \cite{ho17}. This suggests that the disk radius is smaller than the distance between pulsar and companion star at the periastron. In the case of the binary system PSR B1259-63 the decretion disk extends up to the distance of the pulsar periastron. In such a case, the gamma-ray emission above $\sim$ TeV energy range, can be also produced in the comptonization of the disk radiation \cite{van12,boe14}. If this occurs also for
the considered binary system, then the steady, extended TeV emission from the HEGRA TeV $\gamma$-ray source (TeV J2032+4130) can be overcome by the point-like emission from the binary system also above a few TeV during the periastron passage of the pulsar.

\ack
We would like to thank the Referees for many useful comments and suggestions.
This work is supported by the grant through the Polish Naro\-do\-we Centrum Nauki No. 2014/15/B/ST9/04043

\end{document}